\documentclass[twocolumn,showkeys,prd,nofootinbib,floatfix,preprintnumbers,superscriptaddress]{revtex4-1}
\usepackage{multirow}
\usepackage{array}
\usepackage{hyperref}
\usepackage[normalem]{ulem}
\usepackage[utf8]{inputenc}
\usepackage{amsfonts,amsmath,amssymb} 
\usepackage{graphicx,graphics,color}

\usepackage[dvipsnames]{xcolor}
\usepackage{gensymb}
\usepackage{longtable}
\usepackage{bbding}
\usepackage{subfigure}
\usepackage{multirow}
\usepackage{float}
\usepackage{ulem}
\usepackage{soul}
\usepackage{verbatim}
\usepackage{mathtools}

\begin{document}

\title{ Internal symmetry to the rescue: well-posed 1+1 evolution of self-interacting vector fields}

\author{Gabriel G\'omez}
\email{luis.gomezd@umayor.cl}
\affiliation{Centro Multidisciplinario de F\'isica, Vicerrector\'ia de Investigaci\'on, Universidad Mayor\\ Camino La Pir\'amide 5750,  Huechuraba, 8580745, Santiago, Chile\\}

\author{Jos\'e F. Rodr\'iguez}
\email{jose.rodriguez2@correo.uis.edu.co}
\affiliation{Escuela de F\'{\i}sica, Universidad Industrial de Santander,  Ciudad Universitaria, Bucaramanga 680002, Colombia}
\affiliation{ICRANet, Piazza della Repubblica 10, 65122, Pescara PE, Italy}

\begin{abstract}
Previous studies have identified potential instabilities in
self-interacting vector theories associated with the breakdown of the well-posedness of the initial-value problem. However, these conclusions are restricted to Abelian vector fields, leaving room to explore alternative setups, such as non-Abelian vector fields with internal symmetries. Building on this idea, we study the well-posed 1+1 evolution of self-interacting SU(2) vector fields minimally coupled to gravity within the framework of the 't Hooft-Polyakov magnetic monopole configuration. In this context, we present a counterexample in which self-interacting vector fields retain a well-posed initial value problem formulation. Remarkably, this system exhibits the same characteristic speeds as those found in general relativity (GR) in one spatial dimension. Unlike its Abelian counterpart, we achieve
stable numerical evolutions across a wide range of initial conditions within a fully non-linear dynamical background, as evidenced in our time integration algorithm. Although our conclusions are strictly valid for the spherical symmetry case with only magnetic part for the vector field, this study serves as a valuable diagnostic tool for investigating more realistic astrophysical scenarios in three-dimensional settings and under more general background and vector field configurations.
\end{abstract}

\maketitle

\section{Introduction}\label{sec:introduction}

Massive spin-1 (Proca) fields play a central role in both particle physics and cosmology, serving as essential components in the description of fundamental interactions and the evolution of the universe \cite{Peskin:1995ev,Schwartz:2014sze,Weinberg:1996kr}. In particle physics \cite{Grosse-Knetter:1993dzj,Fuchs:1997jv}, vector fields are commonly associated with force carriers gauge bosons that mediate the fundamental forces, such as the electromagnetic, weak, and strong interactions. The dynamics of these vector fields are governed by field equations that describe the propagation and interaction of these forces, such as the Proca equation for massive vector fields \cite{Proca:1936fbw}. These theories extend our understanding of interactions beyond the standard model, often incorporating symmetries and self-interactions that give rise to rich phenomena \cite{Peskin:1995ev,Schwartz:2014sze,Weinberg:1996kr}. In cosmology \cite{Koivisto:2008xf,Esposito-Farese:2009wbc}, vector fields are increasingly explored as candidates for describing cosmic acceleration \cite{
Armendariz-Picon:2004say,Koivisto:2008xf,Gomez:2020sfz,Guarnizo:2020pkj,Gomez:2022okq}, dark matter \cite{Holdom:1985ag,Graham:2015rva,Agrawal:2018vin,Goodsell:2009xc,Bastero-Gil:2018uel,Ema:2019yrd,Kolb:2020fwh,Ahmed:2020fhc,Caputo:2021eaa,Antypas:2022asj} and inflationary dynamics  \cite{Dimopoulos:2006ms,Golovnev:2008cf,Maleknejad:2011jw,Maleknejad:2011sq,Adshead:2012kp,Emami:2016ldl,Garnica:2021fuu}. The potential for vector fields to exhibit self-interactions opens up new avenues for studying exotic compact objects \cite{Minamitsuji:2018kof,Herdeiro:2020jzx,Martinez:2022wsy,Aoki:2022mdn,Bezares:2024btu}, 
and superradiant instabilities in astrophysical systems \cite{Zilhao:2015tya,Baryakhtar:2017ngi,East:2017mrj,East:2017ovw,Wang:2022hra}, among other intriguing phenomena.

Despite their growing popularity, recent studies—both analytical and numerical—have shown that self-interacting vector fields within Einstein’s
gravity are prone to pathologies such as ghost and tachyonic instabilities \cite{Coates:2022qia,Mou:2022hqb,Coates:2023dmz,Coates:2022nif}. In particular, the ghost instability, which arises due to a change in the signature of the effective metric, has been observed dynamically in numerical simulations on a Kerr background, manifesting within a finite time and ultimately leading to the breakdown of the simulation \cite{Clough:2022ygm}.

These issues were also identified in the generalized Proca theory \cite{Tasinato:2014eka,Heisenberg:2014rta,Allys:2015sht,BeltranJimenez:2016rff,GallegoCadavid:2019zke}, a vector-tensor formulation inspired by Horndeski theory \cite{Horndeski:1974wa,Deffayet:2011gz}. Thus, it seems that this issue is not exclusive to Proca fields alone, it also arises in generalizations of the Proca theory \cite{Garcia-Saenz:2021uyv,Unluturk:2023qgk}. 
These findings pose a challenge to the effective field theory description of massive vector fields, particularly those based on effective description of interactions, including dark matter scenarios \cite{Holdom:1985ag,Graham:2015rva,Agrawal:2018vin,Goodsell:2009xc,Bastero-Gil:2018uel,Ema:2019yrd,Kolb:2020fwh,Ahmed:2020fhc,Caputo:2021eaa,Antypas:2022asj}. A potential resolution to this issue is that the effective field theory description breaks down before the ghost instability develops \cite{Aoki:2022woy}. Another approach consists in \textit{fixing the equations} \cite{Barausse:2022rvg,Rubio:2024ryv}, a procedure that ``corrects'' the hyperbolicity character of the system \cite{Cayuso:2017iqc}.  See also Ref.~\cite{Coates:2023swo} for a discussion on the shortcomings of this approach. 

At this point, it is crucial to distinguish between instabilities and the well-posedness of the initial value problem (IVP) (see, e.g., the discussion in Section 1.4 of \cite{Ripley:2020yrx} and Ref.~\cite{Reall:2014pwa}). Cosmological instabilities, particularly gradient instabilities, typically refer to scenarios where linear perturbations grow exponentially fast away from the background solution, leading to unphysical solutions. Yet, identifying such instabilities with the breakdown of the well-posed problem is misleading, as the latter issue is more fundamental \cite{Reall:2014pwa}: it represents a loss of predictivity, meaning that small changes in the initial data could lead to wildly different solutions, or solutions might not exist at all, see also \cite{alcubierre2008introduction}. In a strict mathematical sense, a system of partial differential equations (PDEs) is well-posed if it admits a unique solution that depends continuously on initial data \cite{hadamard,Reula:1998ty}. This property is closely tied to hyperbolicity, which is determined by the principal symbol of the PDE system, a matrix whose elements correspond to the coefficients of the highest derivative terms \cite{Sarbach:2012pr}. Establishing well-posedness requires examining the solution’s behaviour in the high-frequency limit. If the principal symbol has real eigenvalues,  also known as characteristic speeds, and a complete set of eigenvectors, the system is strongly hyperbolic and ensures a well-posed IVP \cite{Hilditch:2013sba}. Based on this distinction, we focus our discussion on the well-posedness of the formulation, rather than the specific instabilities that may arise within a given physical setting. 

It is both legitimate and natural to question whether the issues associated with self-interacting Proca fields are intrinsic to vector fields in general and extend to other vector theories, including those with internal symmetries and non-Abelian structures. Inspired by this line of inquiry, we take a fundamental approach and ask:  
What if the Proca field is endowed with, for instance, an internal \rm{SU(2)} global symmetry? Do the previously identified pathologies persist in this scenario? Through careful investigation, we conclude: \textit{the celebrated problems associated with self-interacting Proca fields do not manifest in the case of non-Abelian SU(2) Proca.} This result not only challenges the implications of previous findings but also underscores the importance of internal symmetries in the description of fundamental fields (see, e.g., \cite{Hoffmann:2024hbh} for a current discussion on self-gravitating bosonic fields with internal symmetries), such as spin-1 fields, within the framework of classical field theories.  

To be more specific, in this work, we assert that the self-interacting non-Abelian SU(2) Proca theory preserves its hyperbolic structure, ensuring the well-posedness of the IVP.
We validate our findings by showcasing the well-behaved 1+1 evolutions of the field variables in a dynamic background. These dynamics exhibit smooth transitions and maintain finite, stable values throughout the time evolution.

At this point, the reader may wonder why we focus specifically on non-Abelian SU(2) fields. The motivation stems from their potential to drive various phenomena in both cosmology \cite{Maleknejad:2011jw,Maleknejad:2011sq,Adshead:2012kp,Garnica:2021fuu,Guarnizo:2020pkj,Garcia-Serna:2025dhk} and astrophysics \cite{Bartnik:1988am,Bizon:1990sr,Greene:1992fw,Dzhunushaliev:2019uft, daRocha:2020jdj,Jain:2022kwq,Martinez:2022wsy,Gomez:2023wei,Martinez:2024gsj}. From a physical perspective, it is well established that non-Abelian interactions exhibit a richer and more intricate phenomenology compared to their Abelian counterparts. One striking difference lies in the behaviour of their mediators—while non-Abelian gauge bosons, such as those in the nuclear forces, can interact with one another, Abelian interactions like electromagnetism lack this self-interaction property \cite{Yang:1954ek,Peskin:1995ev,Schwartz:2014sze,Weinberg:1996kr,tHooft:2005hbu}. This fundamental distinction makes it both challenging and intriguing to uncover novel phenomena that arise uniquely from the non-Abelian nature of vector fields.

This paper follows a systematic structure to present our main findings. After outlining the problem and the main premises in Section \ref{sec:introduction}, we introduce the theoretical framework in Section \ref{sec:theory}. Subsequently, we derive the key equations to be evolved numerically in Section \ref{sec:evol_equations}. Prior to entering the numerical domain, we conduct an analytical examination of the strong hyperbolicity of our theory. We then proceed to numerical simulations in Section \ref{sec:numerical}, employing a broad range of initial data and a detailed exploration of the parameter space of the theory. These simulations enable us to draw compelling conclusions and engage in a thorough discussion presented in Section \ref{sec:conclusions}. In this article, we adopt geometric units, setting $c = G = 1$, where $c$ denotes the speed of light and $G$ represents the universal gravitational constant. Greek indices correspond to space-time coordinates, ranging from 0 to 3, while Latin indices denote SU(2) group indices, varying from 1 to 3. We have adopted the Misner, Thorne and Wheeler (MTW) sign convention for the metric \cite{Misner:1973prb}.

\section{The theory}\label{sec:theory}

We consider a non-Abelian extension of the quartic vector self-interaction $(A_{\mu} A^{\mu})^{4}$, where $A_{\mu}$ represents the Proca field. The theory is described by the following action:
\begin{multline}
    S = \frac{1}{16\pi}\int \sqrt{-g}\, d^4x[R - F_{a\mu\nu}F^{a\mu\nu} -2\mu^{2} B^{\mu}_{a}B_{\mu}^{a} \\
    +\chi_1 B_{a \mu}B^{a\mu}B_{b\nu}B^{b\nu} + \chi_2 B_{a\mu}B^{a}_{\nu}B_{b}^{\mu} B^{b \nu}],\label{eqn:action}
\end{multline}
where $R$ denotes the Ricci scalar, $B_{\mu}^{a}$ is the vector field that belongs to the Lie algebra of the SU(2) group, $F_{a\mu\nu}= \nabla_\mu B_{a\nu}-\nabla_\nu B_{a\mu}+ \tilde{g} \epsilon_{abc}B^{b}{}_{\mu} B^{c}{}_{\nu}$ is the strength tensor, $\tilde{g}$ is the gauge coupling constant, $\epsilon_{abc}$ is the structure constant tensor of the SU(2) group, and $\mu=m_B/\hbar$, with $m_B$ denoting the mass of the vector fields, all of which share the same value. The parameters $\chi_{1}$
and $\chi_{2}$ quantify the strength of the self-interaction with their corresponding terms being linearly independent. This additional scalar invariant, absent in the Abelian theory, emerges from the possible contraction of the spacetime index with the SU(2) index. 

It is worth noting that the Yang-Mills theory is inherently self-interacting. Specifically, it exhibits two types of self-interactions: quartic self-interactions of the form
$ \tilde{g}^2 \epsilon_{abc} \epsilon^{ade} A_{\mu}^{b} A_{\nu}^{c} A^{\mu}_{d} A^{\nu}_{e}$,
and derivative self-interactions of the form
$\tilde{g} \epsilon^{abc} (\nabla_\mu A_{a\nu}) A^\mu_b A^\nu_c$.
The Einstein-Yang-Mills field equations were first numerically evolved in the pioneering work of Choptuik, Chmaj, and Bizon \cite{Choptuik:1996yg}. In their study, they investigated the spherically symmetric collapse of Yang-Mills fields. This work represents a significant example of a well-posed problem involving multiple self-interacting vector fields, providing valuable insights into their numerical time evolution.

As the theory described in \eqref{eqn:action} enjoys, by construction, a global SU(2) internal symmetry, it is convenient to use a vector representation $B_{\mu}^{a}$ for the gauge fields, with a dimension of $3,$ given that the group SU(2) is homomorphic to group SO(3) \cite{arfken2011mathematical}. 

Indeed, this theory is a subset of a comprehensive framework known as the Generalized SU(2) Proca theory \cite{Allys:2015sht,GallegoCadavid:2020dho}, where the internal gauge invariance is replaced by a global invariance. This modification permits the inclusion of additional terms in the action, extending beyond those present in the canonical (massive) Yang-Mills theory \cite{Boulware:1970zc,Shizuya:1975ek,Banerjee:1997sf,tHooft:2005hbu}. This is the spirit behind Horndeski-like theories \cite{Horndeski:1974wa}. The sector described by Eq.~(\ref{eqn:action}) includes only gauge-invariant contractions of the gauge field with itself. Other parts of the theory encompass derivative self-interactions and explicit non-minimal couplings to gravity.

Introducing the gauge covariant derivative $D_{\alpha} F_{a\mu }^{\alpha }  \equiv \nabla_{\alpha} F_{a\mu }^{\alpha } + \tilde{g} B^{b\alpha } \epsilon_{abc} F^{c}_{\alpha\mu}$, the equation of motion can be written as
\begin{equation}
       D_{\alpha }F_{a\mu }^{\alpha } = \mu^2 B_{a\mu } - J_{a\mu} \,,\label{eqn:eqmotion}
\end{equation}
with $J_{a\mu}=\chi_{1} B_{a\mu} B_{b\nu} B^{b\nu} +\chi_{2} B_{a}^{\alpha} B^{b}_{\mu} B_{b\alpha}$. Due to the symmetry of the field strength tensor, the condition $D^{\mu}D_{\alpha }F_{a\mu }^{\alpha }=0$ holds. Consequently, one can properly define conserved charges\footnote{There are also topological charges in this theory \cite{Martinez:2022wsy,Gomez:2023wei}.} in this theory since the action remains invariant under the SU(2) group of global transformations 
\cite{Fuchs:1997jv,Ramond:2010zz}. The structure of the physical charges ultimately depends on the vector field configuration. For the chosen profile in this work, these charges trivially vanish, as demonstrated in Appendix \ref{append:A}.

Applying the gauge covariant derivative to the equation of motion leads to the (generalized) Lorenz condition
\begin{equation}
    \nabla^{\mu}\left(z_{ab} B^{b}_{\mu} \right)=0\,,\label{eqn:lorentzgen}
\end{equation}
where $z_{ab}\equiv \bar{\mu}^{2}g_{ab} + 2\chi_{2} X_{ab}$, $\bar{\mu}^{2}\equiv \mu^{2} + 2 \chi_{1} X$ is the dynamical mass, $X\equiv -\frac{1}{2}B_{c\nu}B^{c\nu}$,  $X_{ab}\equiv -\frac{1}{2}B_{a}^{\mu}B_{b\mu}$ and $g_{ab}=\delta_{ab}$. The Lorenz condition is generally not satisfied for an arbitrary profile. However, as shown in Appendix \ref{sec:Dyonmag}, the specific configuration employed in this work always satisfies this condition.

If $\chi_{2}=0$, then
\begin{equation}
    \nabla^{\mu}\left(\hat{z} B_{a\mu} \right)=0\,,\label{eqn:lorentzred}
\end{equation}
where $\hat{z}\equiv 1+2\chi_{1} X/\mu^2$. Even thought Eq.~\eqref{eqn:lorentzred} exhibits the same structure as the Abelian case (see e.g., \cite{Coates:2022qia}), the corresponding equation of motion does not. Thus, setting $\chi_{2}=0$ does not properly lead to the embedded Abelian solution.

In general, the equation of motion Eq.~\eqref{eqn:eqmotion} can be written in the reduced form
\begin{equation}
     D_{\alpha }F_{a\mu }^{\alpha}=z_{ab} B_{\mu}^{b}\,.
\end{equation}
Recasting the theory in this alternative form highlights that this self-interacting vector theory exhibits a more intricate structure compared to the Abelian case (see, e.g., \cite{Coates:2022qia}), primarily due to the self-interaction term $\chi_{2}$ and the non-Abelian nature of the vector fields. This, by no means, implies that the inclusion of the global SU(2) internal symmetry exacerbates the pathologies of self-interacting vector fields—on the contrary, as we will see, it leads to a well-posed formulation of the IVP.

\subsection{Spherical symmetry case: the 't Hooft-Polyakov configuration}\label{subsec:vector_profile}

We focus on the spherically symmetric case, which restricts the form of both the spacetime metric and the gauge field configuration. To begin, we employ the most general spherically symmetric configuration for the vector field, known as the Witten \emph{ansatz} \citep{Witten:1976ck, Forgacs:1979zs}. This configuration is compatible with the global SU(2) internal symmetry\footnote{Formally, since the adjoint representation is three-dimensional, the SU(2) symmetry group allows for the generation of this configuration.}:
\begin{multline}
    \mathbf{B} = \frac{\tau^a}{\tilde{g}}\Biggl[A_0 \frac{x_a}{r}dt + A_1 \frac{x_a x_j}{r^2} dx^j + \frac{\phi_1}{r}\biggl(\delta_{aj} - \frac{x_a x_j}{r^2}\biggr) dx^j \\
    -\epsilon_{ajk}x^j\frac{(1+\phi_2)}{r^2}dx^k\Biggr] \,,\label{eqn:witten}
\end{multline}
where $\tau_i = -i\sigma_i/2$ is the anti-Hermitian basis for the SU(2) algebra, with $\sigma_i$ denoting the Pauli matrices. Here, $A_0$, $A_1$, $\phi_1$, and $\phi_2$ are functions of the coordinates $(t, r)$; $x_a$ (or $x_j$) are the spacetime Cartesian coordinates; $\delta_{aj}$ is the Kronecker delta; and $\epsilon_{ajk}$ is the Levi-Civita tensor.

An interesting special case of this configuration is the 't Hooft-Polyakov \emph{ansatz}, or magnetic monopole \cite{tHooft:1974kcl,Polyakov:1974ek}, which arises by setting $A_0 = A_1 = \phi_1 = 0$ and $\phi_2 = w(t, r)$:
\begin{equation}
    \mathbf{B}= -\frac{\tau^a}{\tilde{g}}\biggl[ \epsilon_{ajk}x^k\frac{(1+w(t,r))}{r^2}dx^j\biggr] \,.
\end{equation}
The magnetic monopole \emph{ansatz} can be expressed in polar spherical coordinates $(r, \theta, \phi)$ using the standard coordinate transformation from Cartesian coordinates $x^i$. The explicit form of the \emph{ansatz} is given by:
\begin{align} 
\tilde{g} B^1_{\mu}&= [w(t,r)+1][0 , 0 ,  \sin \phi , \sin \theta \cos \theta  \cos \phi ] \,, \nonumber\\
\tilde{g} B^2_{\mu}&=[w(t,r)+1][0 , 0 ,- \cos \phi , \sin \theta  \cos \theta  \sin \phi] \,, \nonumber\\
\tilde{g} B^3_{\mu}&=[w(t,r)+1] [0 ,0 , 0 , -\sin ^2\theta ] \,.\label{eqn:tHPexplicit}
\end{align}
It is important to  highlight two distinctive features of this configuration: (i) its components are transverse, meaning there are no radial components, and (ii) this configuration cannot be realized in the single Abelian case. Moreover, this \emph{ansatz} leads to a self-consistent set of field equations and naturally satisfies the (generalized) Lorenz condition, Eq.~\eqref{eqn:lorentzgen}. In contrast, in the self-interacting Abelian theory \citep{Coates:2022qia,Clough:2022ygm}, this condition must be explicitly enforced (or verified) at each computational grid point. Furthermore, this magnetic monopole configuration has garnered significant interest in astrophysics, particularly in black hole solutions \citep{Volkov:1989fi, Bizon:1990sr,Greene:1992fw, Kleihaus:1997ic}, in the study of critical collapse \citep{Gundlach:1996je,Bizon:2010mp,Rinne:2013qc,Rinne:2014kka,Maliborski:2017jyf,Kain:2019jeg}, and the description of neutron stars \citep{Martinez:2024gsj} and boson stars \citep{Bartnik:1988am,Jain:2022kwq,Martinez:2022wsy}. See \citep{Volkov:1998cc} for a comprehensive review of the Einstein-Yang-Mills case. 

Another notable configuration is the Dyon (see Appendix \ref{sec:Dyonmag}), characterized by $A_1 = \phi_1 = 0$ and $A_0 \neq 0$, $\phi_2 \neq 0$ (see, e.g., \citep{1975PhRvD..11.2227J}). In the pure Einstein-Yang-Mills theory, the only known static, spherically symmetric, and asymptotically flat dyonic solution is the Reissner-Nordström solution with both electric and magnetic charge \cite{Galtsov:1989ip, Ershov:1990qwn, Bizon:1992pi}. However, to date, no other dyon solution has been found in theories involving non-Abelian SU(2) gauge fields that are neither coupled to additional matter fields nor non-minimally coupled to gravity.
This case will not be discussed further and is left for future work. Furthermore, we will refer to the vector gauge fields $B_{\mu}^{a}$ simply as the vector field $w(t, r)$, since it represents the sole dynamical degree of freedom.

Finally, in geometric units, $w$ is dimensionless and the gauge coupling constant $\tilde{g}$ has dimensions $[\tilde{g}] = [L]^{-1}$. Consequently, we will use $\tilde{g}$ to define dimensionless variables. From this point onward, all quantities will be dimensionless, given by the following transformations:
\begin{align*}
t &\to t \, \tilde{g}, \\
r &\to r \, \tilde{g}, \\
M &\to M\, \tilde{g},\\
\mu &\to \mu / \tilde{g}, \\
\chi_{1,2} &\to \chi_{1,2} / \tilde{g}^2, 
\end{align*}
where $M$ represents a variable denoting mass, such as the Misner-Sharp mass $M_{\rm MS}$ \cite{Misner:1964je,Hayward:1994bu}.

\section{Evolution equations in Polar-areal
coordinates}\label{sec:evol_equations}

For the spacetime, we write the line element in Schwarzschild-like coordinates
\begin{equation}
    ds^2 = -e^{2 A(t,r)}dt^2 + e^{2B(t,r)}dr^2 + r^2 d\Omega^2,\label{eqn:line_element}
\end{equation}
where the metric components depend on $(t,r)$. Considering the aforementioned \textit{ansazts} for the metric \eqref{eqn:line_element}  and for the vector fields Eq.~(\ref{eqn:tHPexplicit}), the field equations can be calculated. The $tt$ and $rr$ components of the Einstein field equations are, respectively,
\begin{multline}
    2 r^2 \left[e^{-2 B} \left(r B'-w'^2\right)-e^{-2 A} \dot{w}^2\right]+\left(1-e^{-2 B}\right) r^2\\
    -\left(w^2-1\right)^2 -2 \mu^2 r^2(w+1)^2+\chi (w+1)^4 =0,
\end{multline}
and
\begin{multline}
    2 r^2 \left[e^{-2 B} \left(r A'-w'^2\right)-e^{-2 A} \dot{w}^2\right]-\left(1-e^{-2 B}\right) r^2\\
    +\left(w^2-1\right)^2+2 \mu^2 r^2 (w+1)^2-\chi(w+1)^4 =0.
\end{multline}
The evolution equation for the vector field is given by
\begin{multline}
   r^2\ddot{w}+r^2 \dot{w} \left(\dot{B}-\dot{A}\right)-r^2e^{2 (A-B)} \left[w' \left(A'-B'\right)+w''\right]\\
   +e^{2 A} \left[w \left(w^2-1\right)+\mu^2 r^2 (w+1)-\chi(w+1)^3 \right]= 0,
\end{multline}
where overdot and a prime denote time and spatial derivatives, respectively.
It is convenient to introduce an effective parameter, since the self-interaction parameters of the theory can be recast as a linear combination in the field equations\footnote{This simplification is only possible under the ’t Hooft–Polyakov monopole configuration, even at the perturbative level. When additional degrees of freedom are present, it no longer holds, and each self-interaction term must be analyzed independently.}: $\chi=2\chi_{1}+\chi_{2}$. This reduction leaves us with only two free parameters: the particle mass $\mu$ and the effective self-interaction parameter $\chi$. This implies that, in the limit $\chi\to0$, or equivalently  $\chi_{2}=-2\chi_{1}$, the present theory reduces to the standard (non-Abelian) Proca theory.

We can always write down the equations of motion as a first-order system defining~\cite{Choptuik:1996yg} 
\begin{equation}
    Q \equiv   w' = \partial_{r} w\;,\;\;\;  P \equiv e^{B - A}\dot{w} = e^{B- A} \partial_{t} w\;.
\end{equation}
From here, the time-evolution equations are give by
\begin{align}
 E_{Q}\equiv    \partial_t Q & - e^{A - B}\partial_r P - \partial_r(A - B)e^{A - B}P=0\;,\label{eqn:evol1} \\ 
E_{P}\equiv    \partial_t{P} & - e^{A - B}\partial_r Q - \partial_r(A - B)e^{A - B}Q - V=0\;,\label{eqn:evol2}\\
E_{w}\equiv     \partial_t w & - e^{A - B} P=0\;,\label{eqn:evol3}
\end{align}
while the constraints equations are
\begin{multline}
 C_{A}\equiv    \partial_r A - \frac{Q^2}{r} - \frac{P^2}{r} + \frac{(1 - e^{2B})}{2r} + \frac{e^{2B}}{2r^3}[(1-w^2)^2  \\ +2\mu^2 r^2(w+1)^2   - \chi(w+1)^4]=0,\label{eqn:constraintA}
\end{multline}
\begin{multline}
C_{B}\equiv    \partial_r B -  \frac{Q^2}{r} - \frac{P^2}{r} - \frac{(1 - e^{2B})}{2r} - \frac{e^{2B}}{2r^3}[(1-w^2)^2 \\ + 2\mu^2 r^2(w+1)^2  - \chi(w+1)^4]=0,\label{eqn:constraintB}
\end{multline}
where we have defined an effective potential as 
\begin{equation}
    V = \frac{e^{A+B}}{r^2} [w(1-w^2) -\mu^2 r^2(1+w)+ \chi(1 + w)^3]\;.\label{eqn:potential}
\end{equation}
We recall that all the variables in the previous equations are dimensionless.

We can recognize that the evolution equations $E_{Q}$ and $E_{P}$ resemble those of GR coupled to a canonical massless scalar field, but with an effective potential given by Eq.~(\ref{eqn:potential}). Consequently, one might expect this theory to also be well-posed, as the potential does not involve derivative terms that could alter the principal part of the system.

To check the hyperbolicity of the system, we follow the standard procedure to compute the characteristic
speeds, as outlined in Refs.~\cite{gustafsson1995time,Sarbach:2012pr,Kovacs:2020ywu,Ripley:2022cdh,courant2008methods}
We begin by expressing the system in the following compact form: $E_{I}(\partial_{t}\textbf{V},\partial_{r}\textbf{V},\textbf{V},\textbf{W})$, where $\textbf{V}=(P,Q)$ and $\textbf{W}=(A,B)$, with the index $I$ labeling the number of equations. Introducing the characteristic covector $\xi_{\mu}$, the principal symbol can be written as
\begin{equation}
    \mathcal{P}_{IJ}(\xi)=\mathcal{P}_{IJ}^{\mu}\xi_{\mu}=\frac{\partial E_{I}}{\partial (\partial_{\mu} \textbf{V}^{J})} \xi_{\mu}\,.
\end{equation}
A necessary condition for the system to be well-posed is obtaining real and distinct solutions for the characteristic equation: $\text{det}(\mathcal{P}_{IJ}(\xi))=0$. In spherical symmetry, the principal symbol matrix is determined by \cite{Ripley:2022cdh}
\begin{equation}
    \mathcal{P}(\xi)=\begin{pmatrix}
\mathcal{A}\xi_{t} + \mathcal{B}\xi_{r}& \mathcal{Q}\xi_{r}\\
\mathcal{R}\xi_{r} & \mathcal{S}\xi_{r} 
\end{pmatrix}\,.
\end{equation}
Exact values of quantities $\mathcal{A}$,$\mathcal{B}$,$\mathcal{Q}$, $\mathcal{R}$ and $\mathcal{S}$ can be found, for instance, in Appendix A of \cite{Franchini:2022ukz,Thaalba:2024htc}. The characteristic speeds are defined as $c\equiv-\frac{\xi_{t}}{\xi_{r}}$ and are given by
\begin{align}
    c_{\pm} = \frac{1}{2} \left(\text{Tr}(\mathcal{C}) \pm \sqrt{\mathcal{D}}\right), \label{eqn:char_speeds}
\end{align}
where, 
\begin{align}
    \mathcal{D} & \coloneqq \text{Tr}(\mathcal{C})^2 - 4\text{Det}(\mathcal{C}), \\ 
    \mathcal{C} & \coloneqq \mathcal{A}^{-1} \cdot\left(\mathcal{B}-\mathcal{Q} \cdot \mathcal{S}^{-1} \cdot \mathcal{R}\right).
\end{align}
For this theory, the characteristic speeds are equivalent to those of GR
\begin{equation}
 c_{\pm}=\pm e^{A-B}\,.   
\end{equation}
Hence, the condition for strong hyperbolicity is fully guaranteed \cite{Hilditch:2013sba}. Having established that our system is strongly hyperbolic, a feature that implies it is also well-posed, we can now proceed to perform numerical simulations with suitable initial data and appropriate boundary conditions.

\subsection{Initial data}\label{sec:numerical}
We choose a localized Gaussian family of initial regular data for the vector field, which we refer to as the Type I family. This is defined as\footnote{This particular family of initial data also exhibits mass scaling in the supercritical regime for the Yang-Mills field, as showed by Choptuik et al. \cite{Choptuik:1996yg}, similar to his seminal work of massless scalar field \cite{Choptuik:1992jv}.} 
\begin{align}
    w(0,r) &=  -1 + a_{0} \exp{\left[-\left(\frac{r-r_{0}}{w_{0}}\right)^{2}\right]},\\
    Q(0,r) &= \partial_{r} w(0,r),\\
    P(0,r) &=0.
\end{align} 
In addition, we consider a time-symmetric kink-like initial data configuration \cite{Choptuik:1996yg,Choptuik:1999gh}, referred to as the Type II family, given by
\begin{align}
      w(0,r) &=  \left[1 + a_{0} \left(1 + \frac{b r}{w_{0}} \right) e^{-2(r/w_{0})^2} \right] \tanh \left(\frac{r_{0} - r}{w_{0}} \right),\\
    Q(0,r) &= \partial_{r} w(0,r),\\
    P(0,r) &=0.
\end{align} 
Here $a_{0}$ is the amplitude, $r_{0}$ is the grid location of the pulse and $w_{0}$ denotes the root-mean-square width of both pulses. In general, we choose the parameters such that $w(0,0)=-1$ and $w^{\prime}(0,0)=0$, ensuring consistency with Eq.~(\ref{eqn:tHPexplicit}). For the Type II family, enforcing these conditions leads to the derived relations: $a_{0}=-1 - \coth{(\frac{r_{0}}{w_{0}})}$ and $b=-1 + \coth{(\frac{r_{0}}{w_{0}})}$. No such conditions are imposed on the Type I family. We have used a standard nomenclature to ease readability and comparison~\footnote{In this sense, we have ambiguously used the same letter, $w$, to describe both the width of the pulse and the (initial) value of the vector field. However, the former is denoted as $w_{0}$, while the latter is written as $w(0,r)$.}.

\subsection{Numerical scheme}\label{sec:numerical}

For this work, we adopt the fully constrained approach and utilize the method of lines for time evolution with a fourth-order Runge–Kutta scheme, with spatial derivatives discretized via second-order finite difference approximation satisfying summation by parts~\cite{Calabrese:2003vx}. For a comprehensive discussion on the numerical implementation and techniques employed, we direct the reader to Refs.~\cite{Franchini:2022ukz,Thaalba:2024htc}.

The regularity conditions at the origin must be satisfied for the metric variables, which we impose as follows:
\begin{equation}
    \partial_{r} A(t,0)=0,\;\; A(t,0)=0,\;\;B(t,0)=0,\;\; \partial_{r} B(t,0)=0,\label{eqn:regcondmet}
\end{equation}
and
\begin{equation}
    P(t,0)=0,\;\;Q(t,0)=0,\label{eqn:regrcondfields}
\end{equation}
as required by the constraints equations. Additionally, we impose the condition $w(t,0)=-1$. 
Furthermore, we impose outgoing Sommerfeld boundary conditions on the $Q$ and $P$ fields at $r_{\rm max}.$ 

Through the numerical evolution, we monitor the constraint equations by computing the cumulative error (\textit{residual}) at every time step for the selected resolutions. 
As demonstrated in Appendix \ref{append:C}, our simulations effectively converge with second-order accuracy as the resolution is increased. 

At this point, we are fully equipped with all the physical and mathematical tools to confidentially explore the dynamical evolution of the vector field.

\subsection{Results}\label{sec:results}

In this section, we present the results of numerical simulations we performed, which enabled us to identify the regime of well-posedness for the self-interacting vector theory in a dynamical regime in spherical symmetry. The dynamical evolution of the vector fields is analyzed within a two-dimensional parameter space ($\mu,\chi$), keeping the initial data parameters fixed unless otherwise specified. Note that, in geometric units, $\chi$ has dimensions of inverse square length.

The parameters for type I initial data are set as follows: $r_{0}=40$, $w_{0}=3$ and $a_{0}=0.5$. This value of the amplitude is nearly at the threshold for inducing gravitational collapse, yet remains safely below it. Meanwhile, for type II initial data, the parameters are set as follows: $a_{0}=-2.10479$, $r_{0}=30$, $w_{0}=20$, $b=0.104791$ and $\chi=1.2$. We begin by adjustinging the particle mass to $\mu=0.01$ and varying the self-interaction parameter over the range $\chi \in [-150,50]$ for type I data and $\chi \in [-50,2]$ for type II data.  These ranges enable us to explore both the weak and strong regimes of self-interaction.

For illustrative purposes, and after exploring one-parameter family, we present a selection of representative cases that encapsulate the main distinctive features of the evolution. 
As a reference, we have included the standard (non-interacting) Proca theory ($\chi = 0$) in our analysis. However, we have excluded the pure Yang-Mills case, as the small-mass limit of the vector field under consideration, along with setting $\chi=0$, shows no significant differences in dynamical evolution compared to the Yang-Mills case, as expected. Therefore, our primary focus is on exploring the effects of the self-interaction parameter on the dynamical evolution of both types of initial data. The main results are summarized as follows:\\
\begin{figure*}[ht!]
\centering
\includegraphics[scale=0.35]{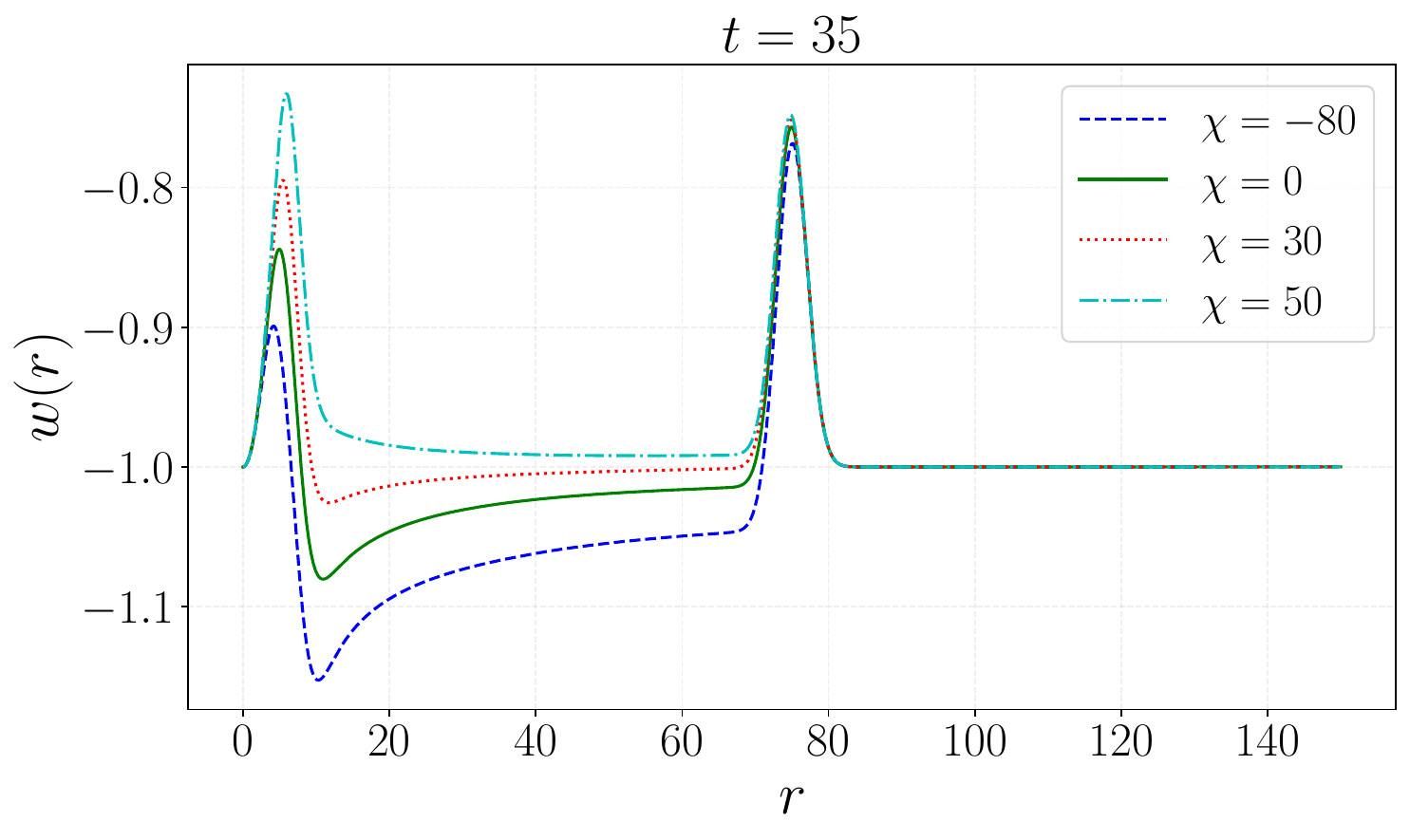} 
\includegraphics[scale=0.35]{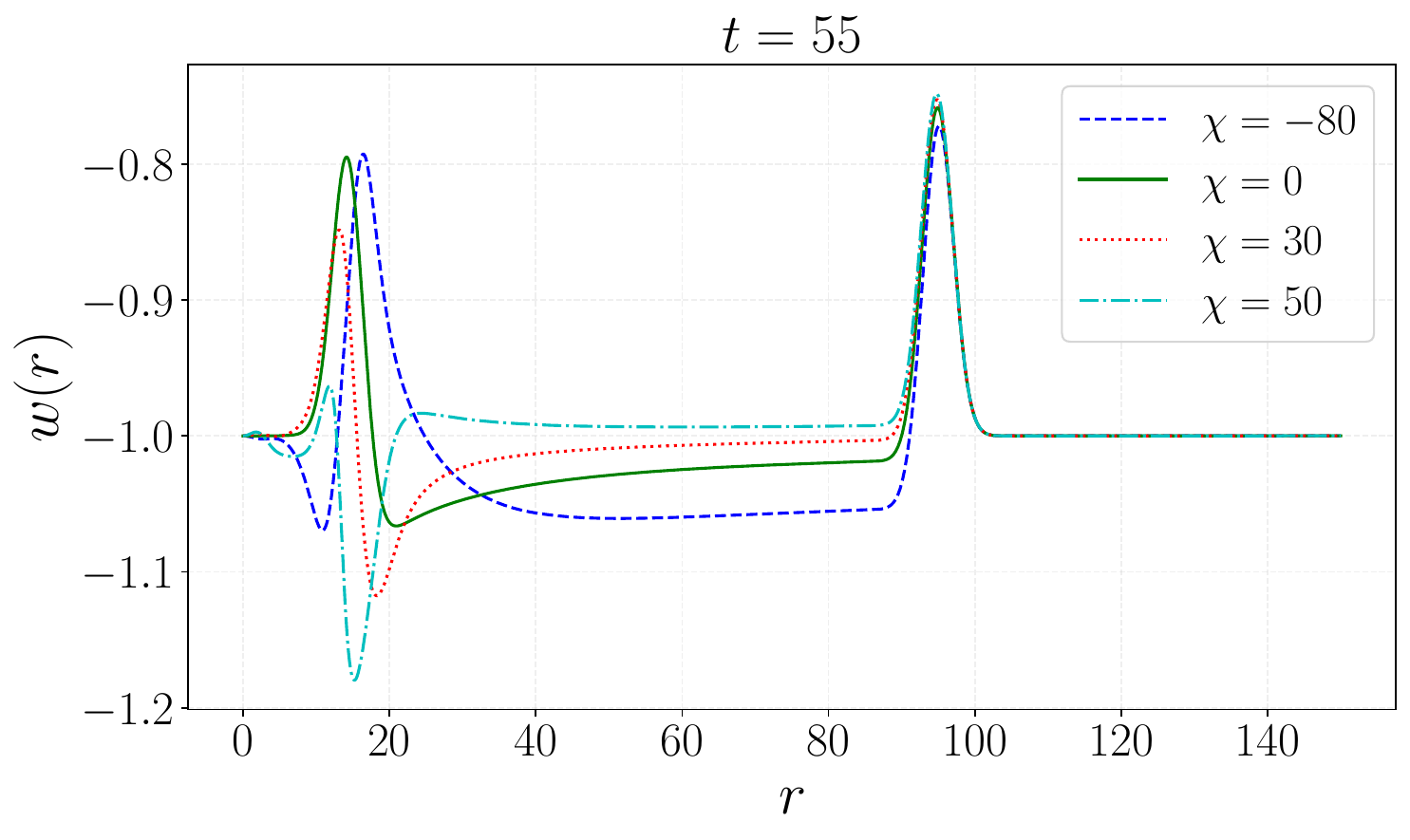} \includegraphics[scale=0.35]{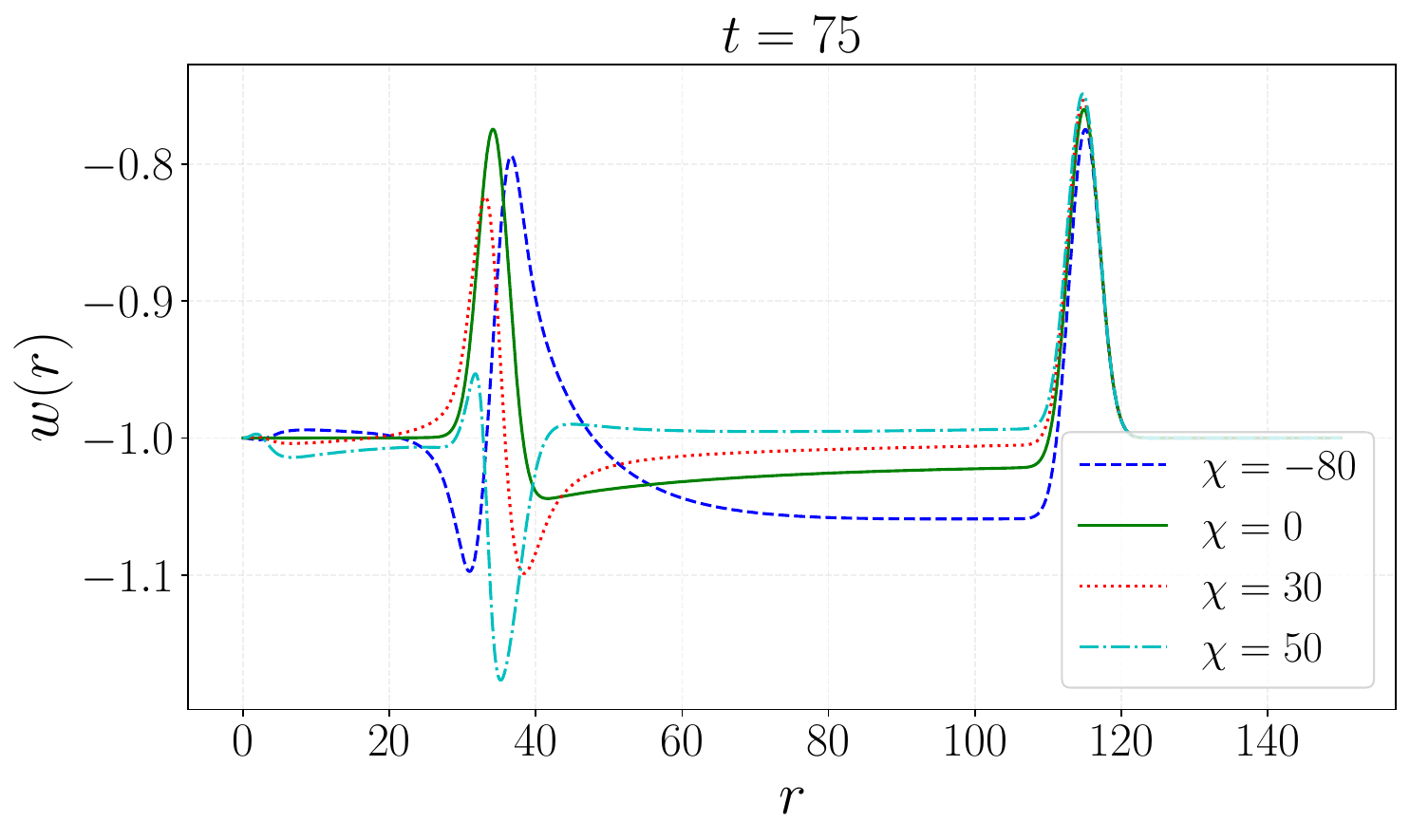}
\includegraphics[scale=0.35]{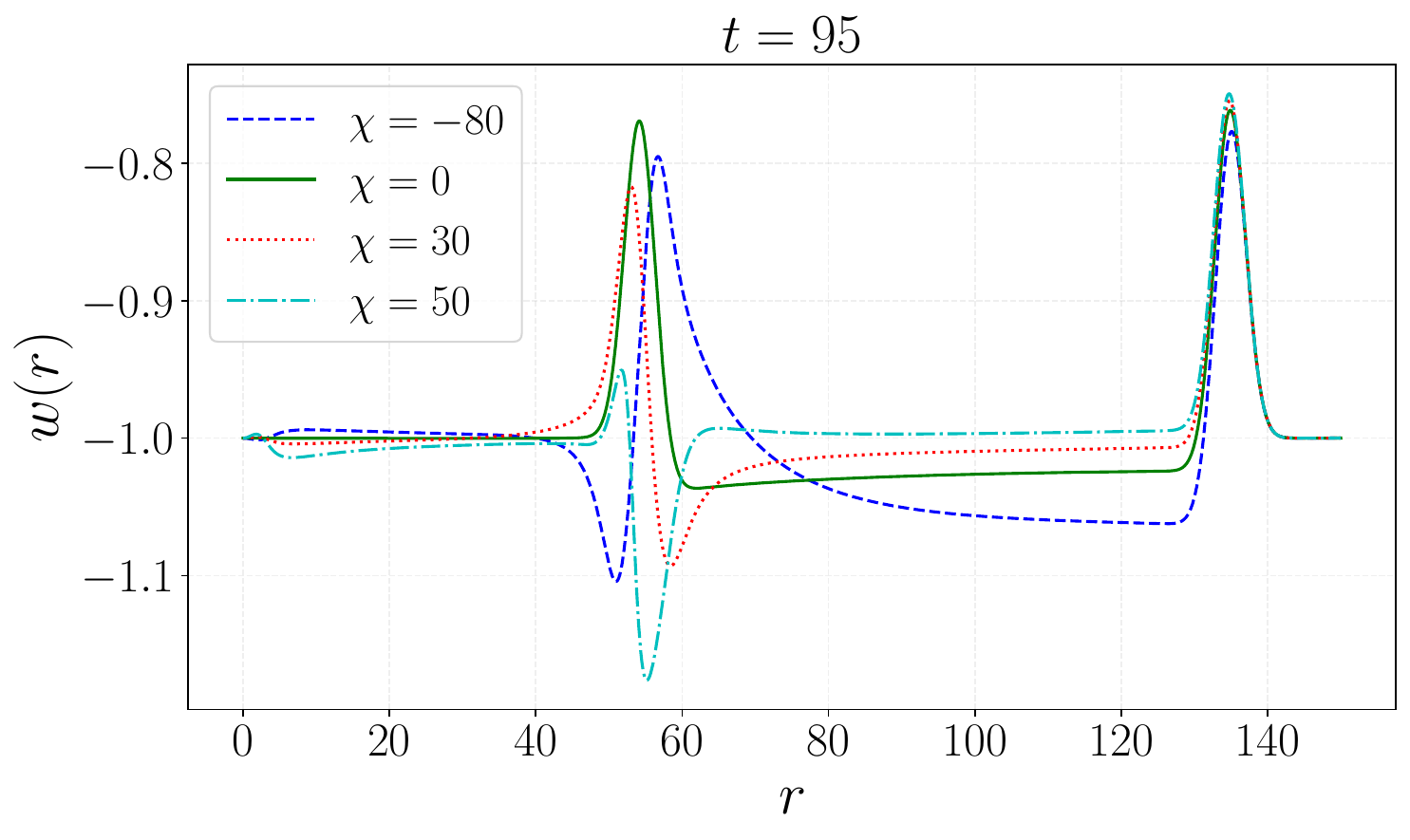} \caption{\textbf{Type I data}: Snapshots of the evolution of the vector field for different values of the self-interaction parameter $\chi$, as indicated in the legend. For large negative values of $\chi$
the field undergoes rapid dispersion before bouncing, with relatively small amplitudes compared to the other cases. For $\chi=0$, the evolution is less dynamic but still exhibits dispersion, with the field maintaining a more stable amplitude throughout the entire evolution. In contrast, for positive values of $\chi$, the field develops significantly larger amplitudes near the origin ($r=0$) and reflects fully, following the periodic pattern shown in the left panel of Fig.~\ref{fig:evol_vector_full}.  This simulation showcases the impact of the self-interaction parameter on the vector field's dynamics, resulting in a redistribution of energy  within the system.}
\label{fig:evol_vector_chi} 	
\end{figure*}

\begin{figure*}[ht!]
\centering
\includegraphics[scale=0.35]{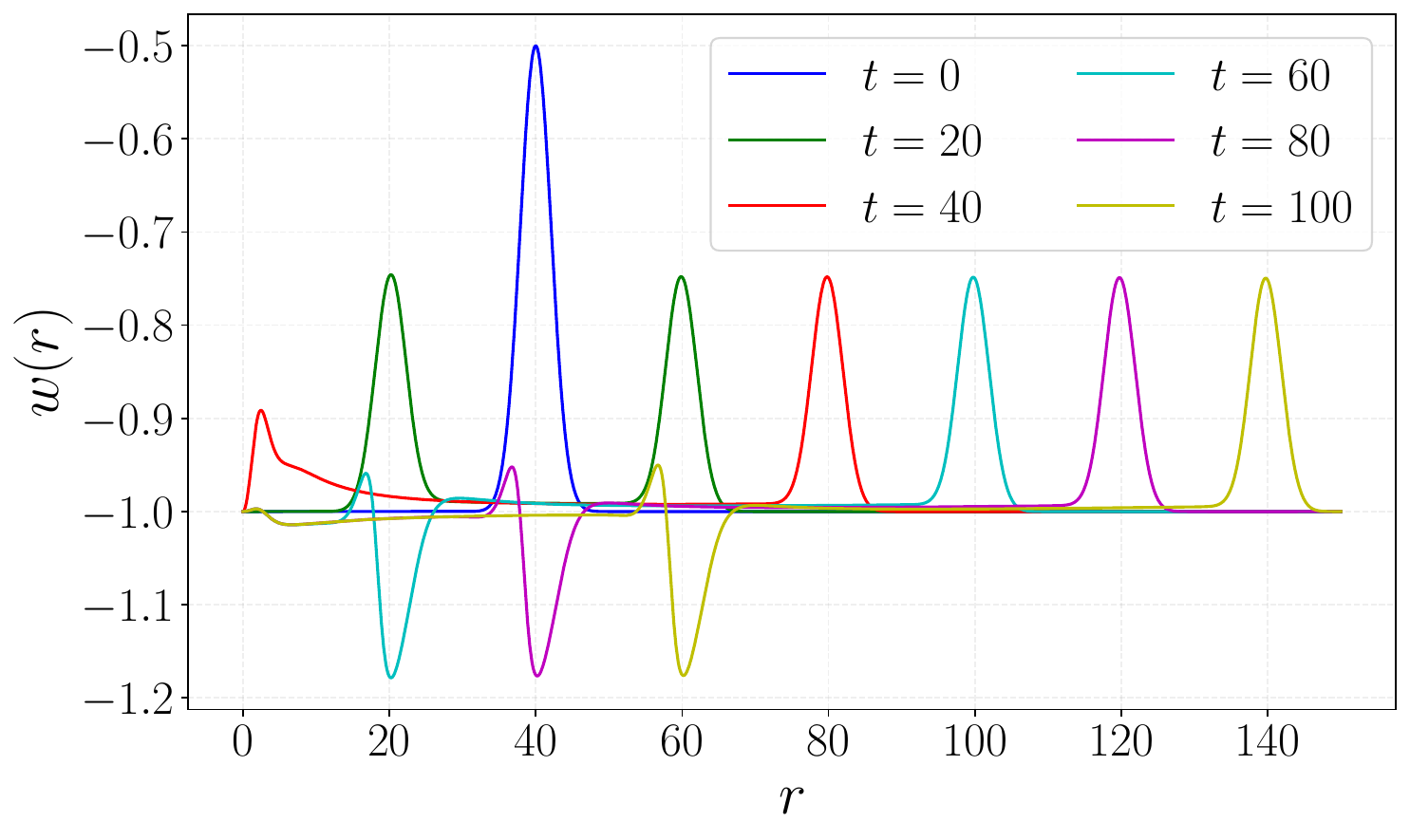} 
\includegraphics[scale=0.35]{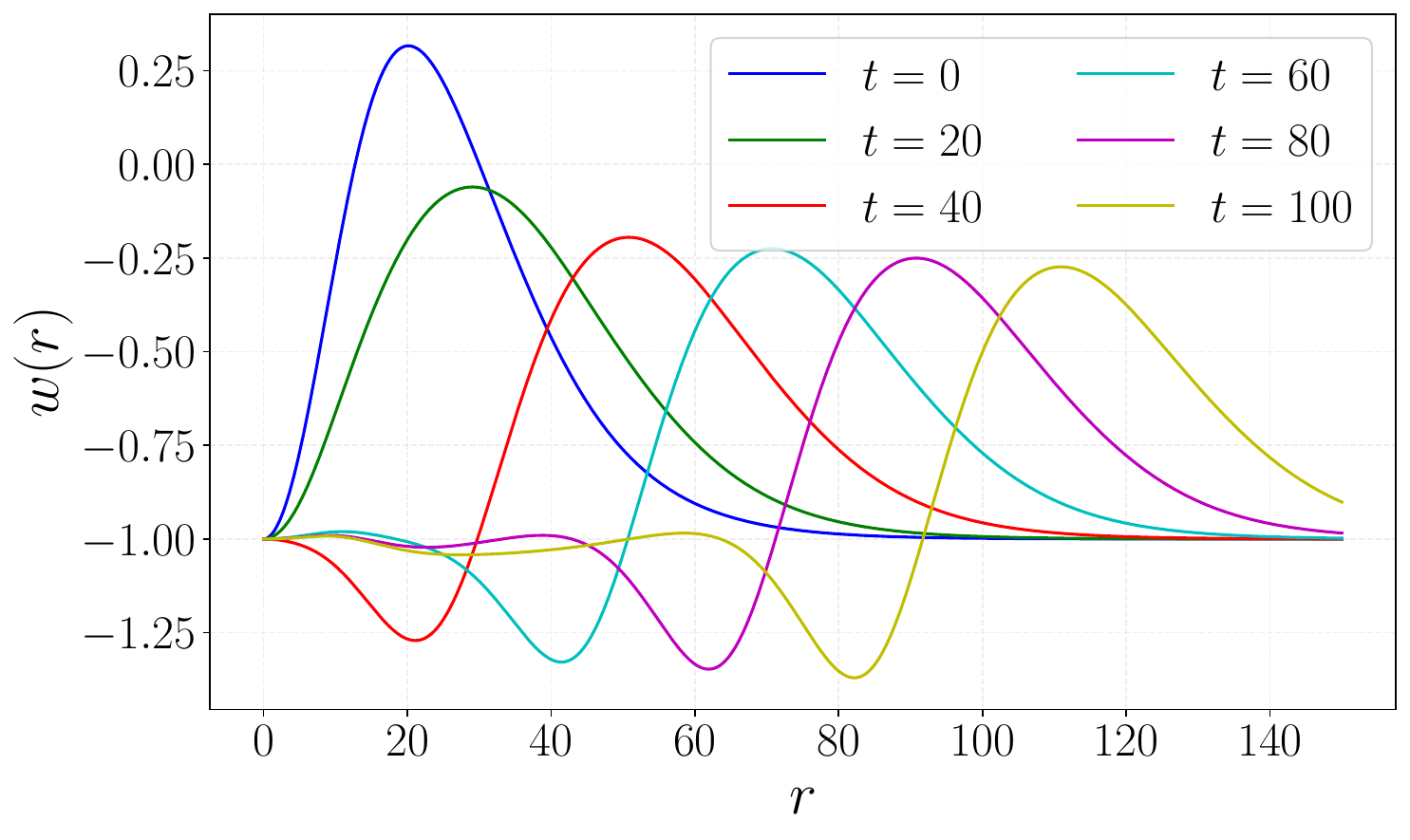} 
\caption{Snapshots of the evolution of the vector field for $\chi=50$ and $\chi=1.2$, corresponding, respectively, to \textbf{Type I initial data} (\textit{left panel}) and \textbf{Type II initial data} (\textit{right panel}). For the type I family, it is evident that the amplitude diminishes as the initial pulse splits into two distinct pulses: one travelling inward (ingoing pulse) and the other propagating outward toward the numerical boundary at $r=150$. For this value of $\chi$, the field fully reflects after bouncing at $r=0$. For the type II family, the outgoing pulse also reduces its amplitude, exhibiting time-symmetric dispersion as it moves outward toward the outer boundary.}
\label{fig:evol_vector_full} 	
\end{figure*}

\textbf{Type I Initial Data:}
\begin{itemize}
    \item As a general trend (see Figs.~\ref{fig:evol_vector_chi} along with the left panel of Fig.~\ref{fig:evol_vector_full}), the pulse splits into two distinct components travelling in opposite directions. The ingoing pulse rapidly decreases in amplitude and reflects at $r = 0$, while the outgoing pulse disperses as it approaches the outer boundary.
    
    \item For $\chi \lesssim -20$, we observe that the ingoing wave packet partially reflects upon rebounding. However, most of the wave packet does not reflect but continues to disperse outward toward infinity. Notably, the amplitude of the reflected pulse increases as $\chi$ takes increasingly negative values.
    
    \item For $\chi = 0$, no reflection is observed, while partial reflection begins again for $\chi \gtrsim 20$, with the amplitude of the reflected pulse increasing significantly for larger values of $\chi$.
    
    \item Interestingly, for $\chi = 50$, the ingoing pulse increases in amplitude just before rerouting (see the cyan curve in the top-left panel of Fig.~\ref{fig:evol_vector_chi}, corresponding to $t = 35$) and then gradually decreases. For even larger values of $\chi$, the system becomes \textit{stiff}. However, this issue is readily circumvented by reducing the amplitude to minor sub-percent level variations, indicating that it is merely an artefact of the initial data rather than an inherent pathology of the theory\footnote{This is not the case, however, for the Abelian theory. It has been shown that even arbitrarily low amplitude initial data can still result in the breakdown of the theory \cite{Coates:2023dmz}.}. Nevertheless, the effect of the amplitude is further explored when we take a closer look at the spherical collapse in Section \ref{sec:collapse}.
    
    \item For small amplitude values $a_{0} < \mathcal{O}(0.1)$, no reflection is observed after bouncing.
\end{itemize}

\begin{figure*}[ht!]
\centering
\includegraphics[scale=0.35]{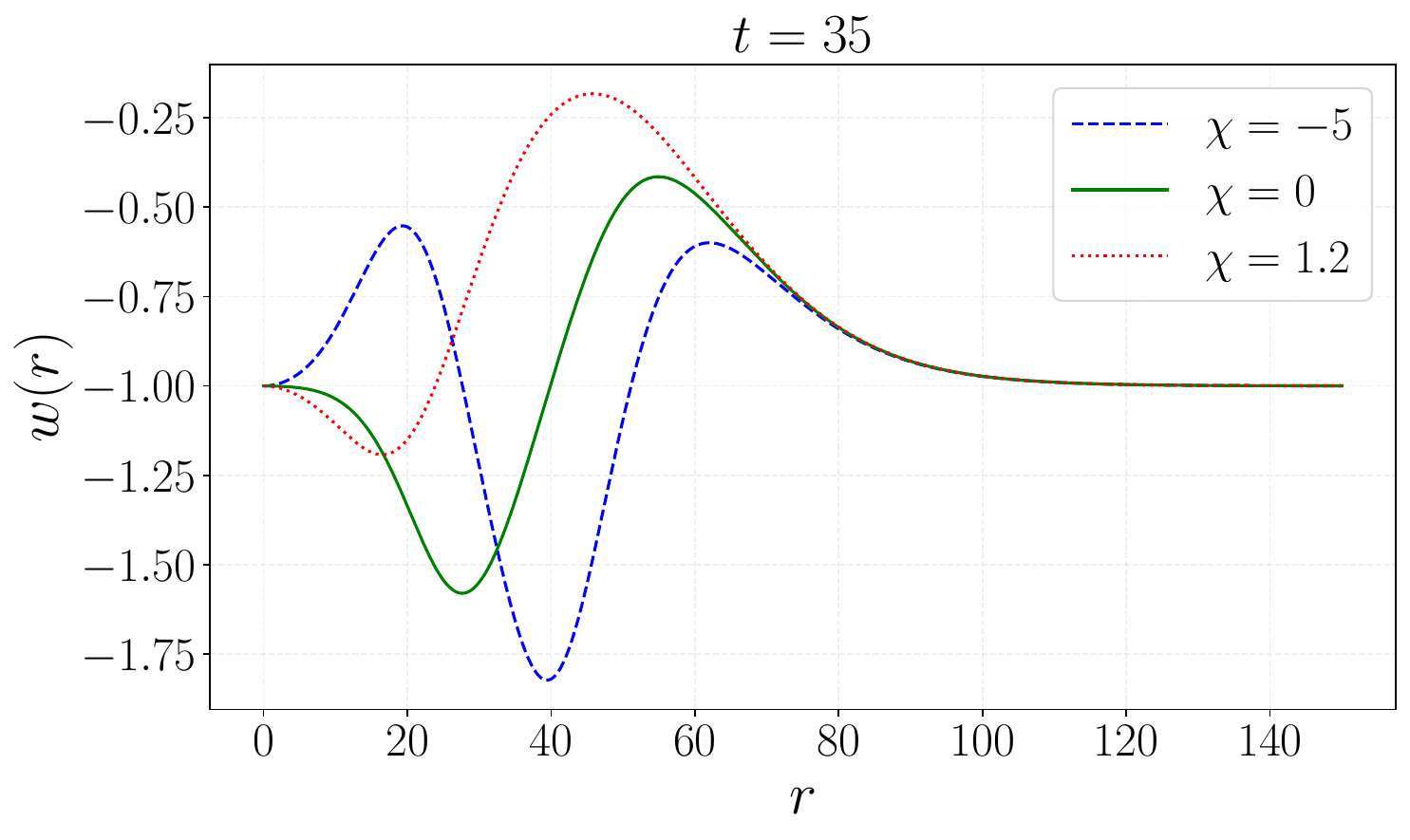} 
\includegraphics[scale=0.35]{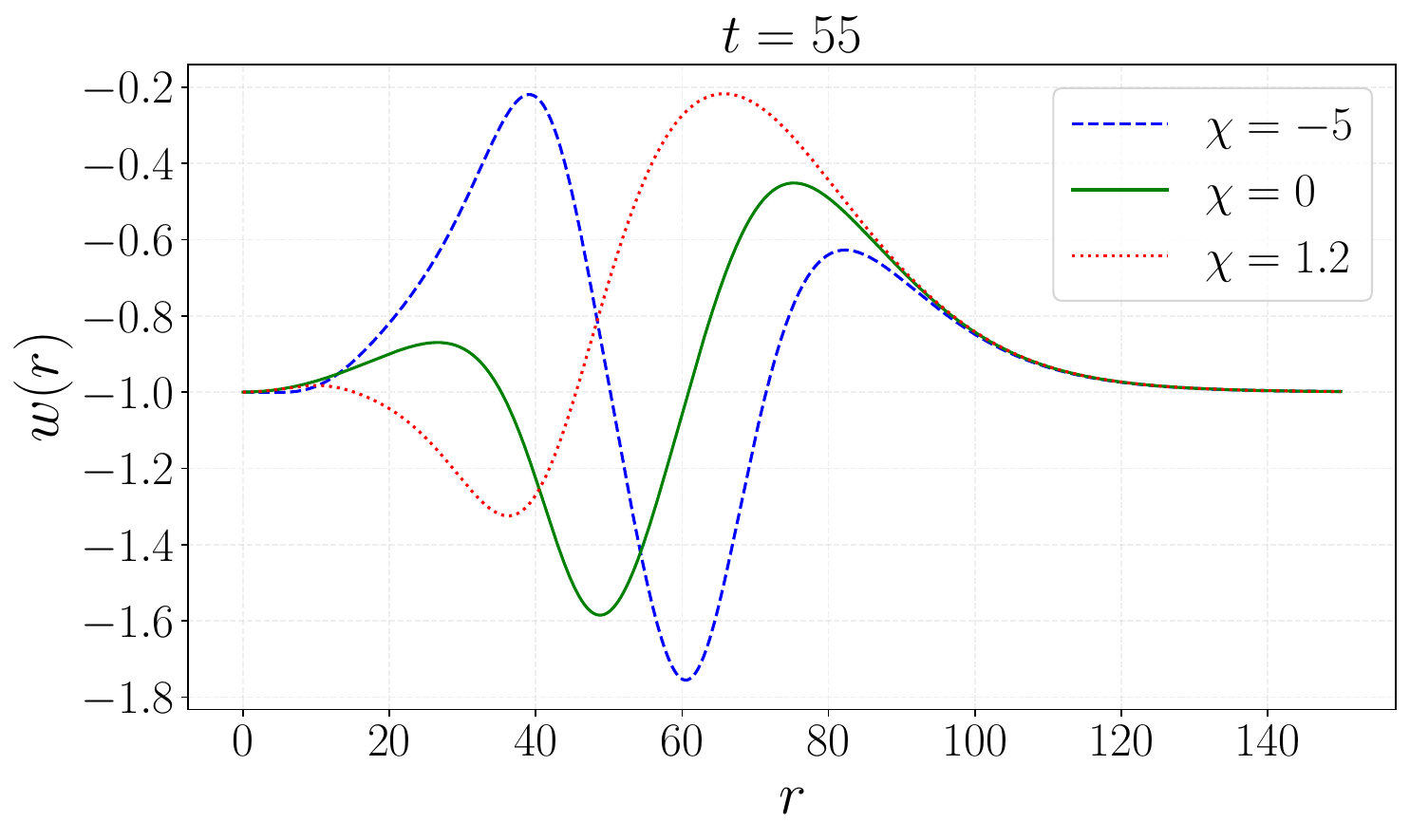} \includegraphics[scale=0.35]{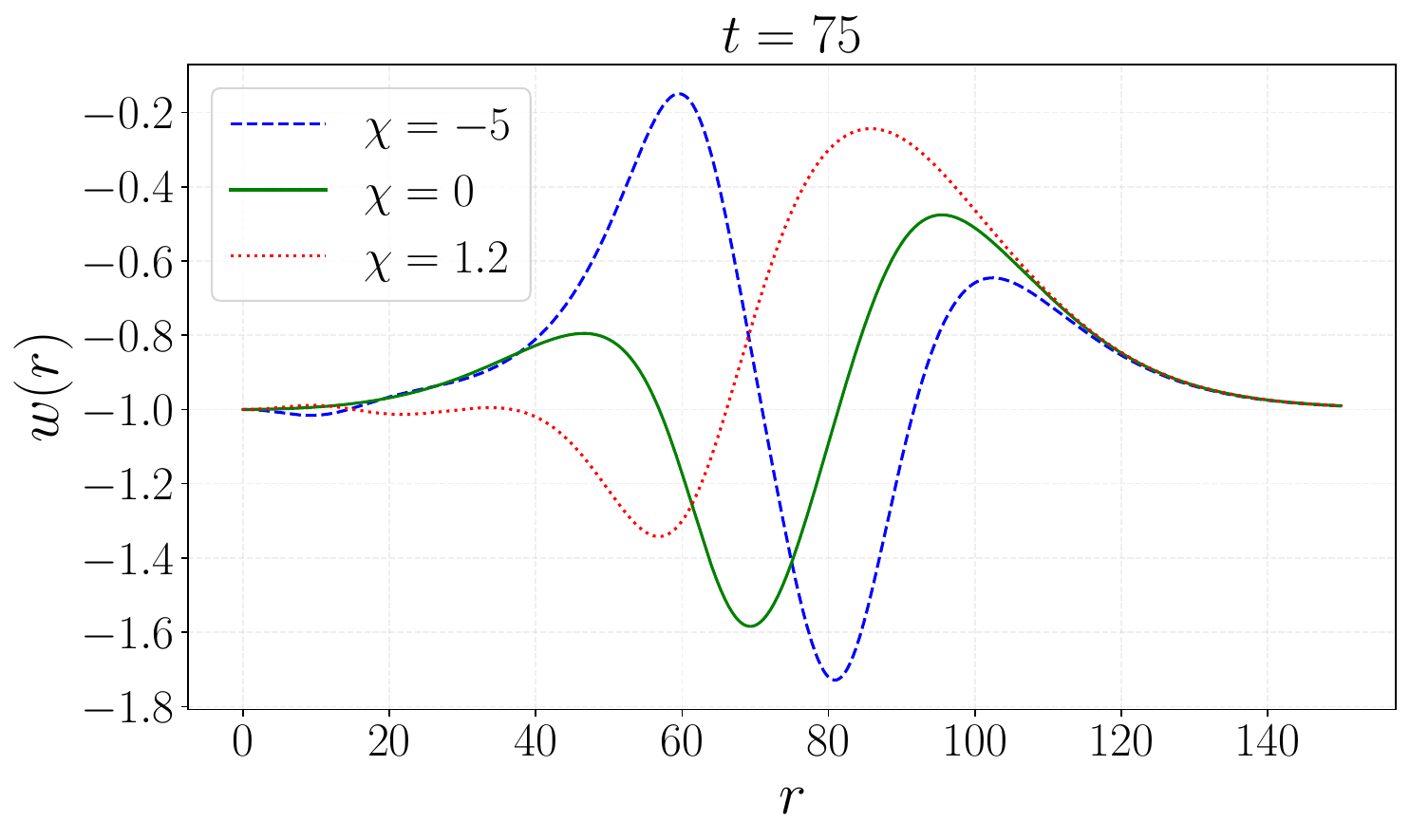}
\includegraphics[scale=0.35]{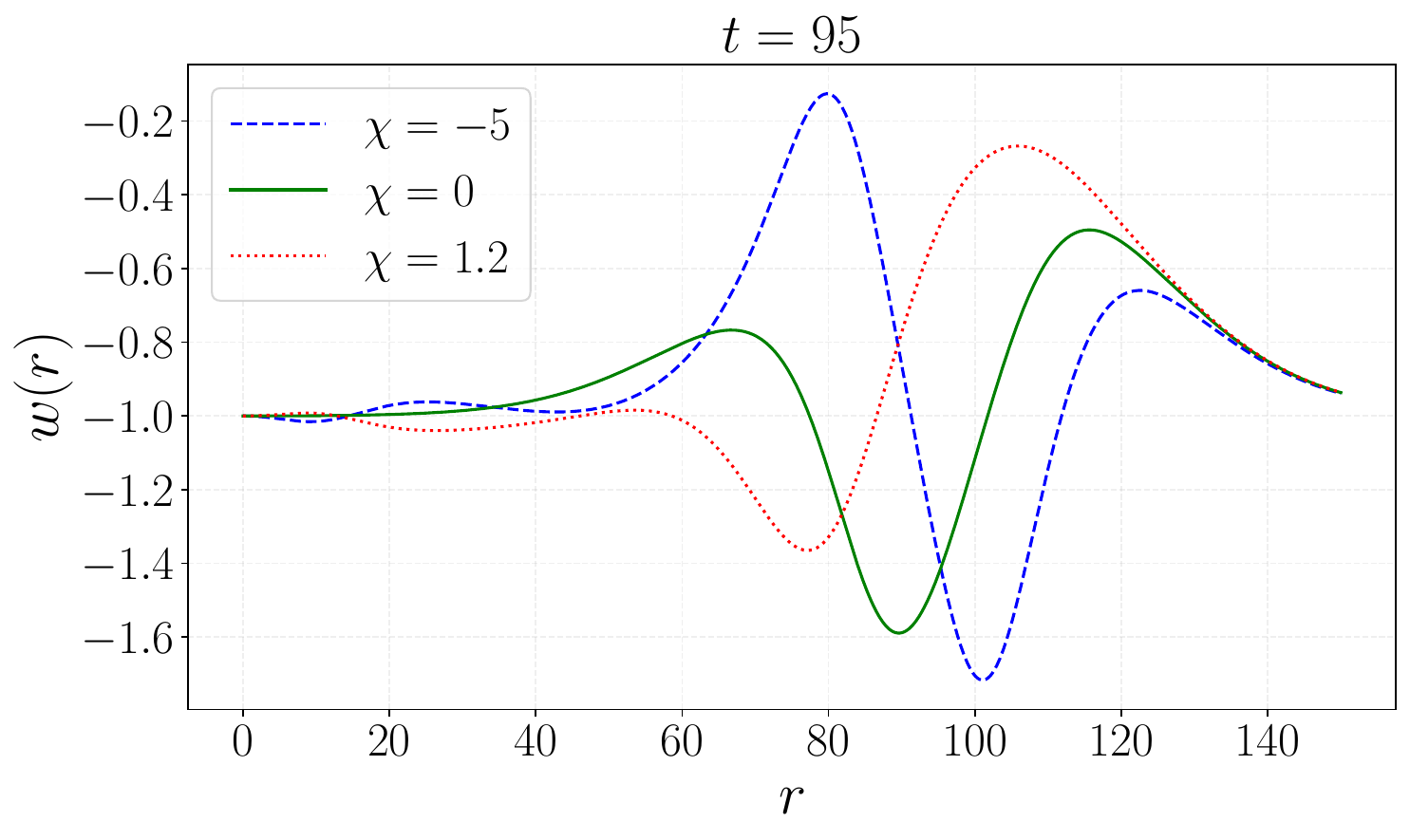} \caption{\textbf{Type II data}: Snapshots of the evolution of the vector field for different values of the self-interaction parameter $\chi$, as indicated in the legend. For negative values of $\chi$
the field undergoes rapid dispersion once it stars evolving, with relatively large amplitudes compared to the other cases. For $\chi=0$, the evolution is less dynamic but still exhibits dispersion, with the field maintaining a more stable amplitude throughout the entire evolution. In contrast, for positive values of $\chi$, the field remains more coherent as it propagates, following the periodic pattern shown in the right panel of Fig.~\ref{fig:evol_vector_full}. This simulation showcases the impact of the self-interaction parameter on the vector field's dynamics, resulting in a redistribution of energy  within the system.}
\label{fig:evol_vector2_chi} 	
\end{figure*}

\textbf{Type II initial data:}
\begin{itemize}
    \item  The behaviour of this initial data follows a general trend as observed in our simulations (see Figs.~\ref{fig:evol_vector2_chi} along with the right panel of Fig.~\ref{fig:evol_vector_full}):  as the outgoing pulse propagates, it disperses in a time-symmetric manner while approaching the outer boundary.
    \item We find that the self-interaction parameter $\chi$ plays a crucial role in shaping the evolution of the outgoing pulse. Specifically, for positive values of $\chi$, the pulse remains more coherent as it propagates, while for negative values, it leads to an increase in amplitude.. This behaviour directly impacts the overall energy distribution of the system. Notably, even a slight increase in $\chi$ by just a few percent near the critical regime can dramatically alter the system’s dynamics, potentially driving the pulse toward collapse. This phenomenon is explored to some extent in Section \ref{sec:collapse}, although it remains a subject of ongoing investigation.
\end{itemize}

To better illustrate the phenomenon of reflection for Type I data, we present the particular case $\chi = 50$ to effectively capture the impact of self-interaction on the wave evolution, an effect absent in the noninteracting scenario. This is shown in the left panel of Fig.~\ref{fig:evol_vector_full}, which provides snapshots that span the full-time evolution of the vector field.

We have also presented the full-time evolution of the vector field for Type II data, setting \(\chi = 1.2\). This is illustrated in the right panel of Fig.~\ref{fig:evol_vector_full}. Unlike Type I data, this profile does not rebound, as it is not split into two. Instead, there is only an outgoing pulse exhibiting time-symmetric dispersion as it moves outward toward the outer boundary. 

\begin{figure*}[ht!]
\centering
 \includegraphics[scale=0.35]{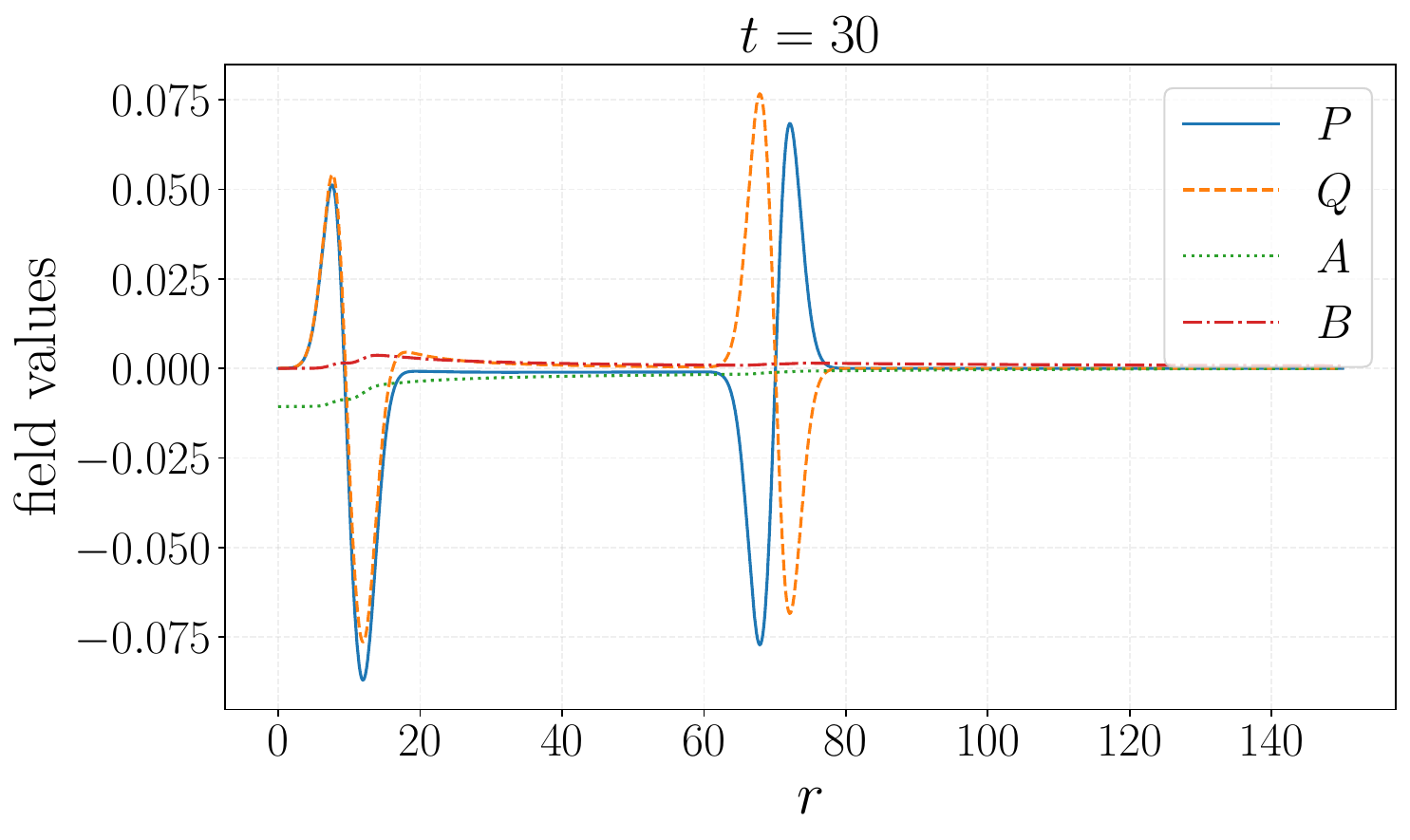}
\includegraphics[scale=0.35]{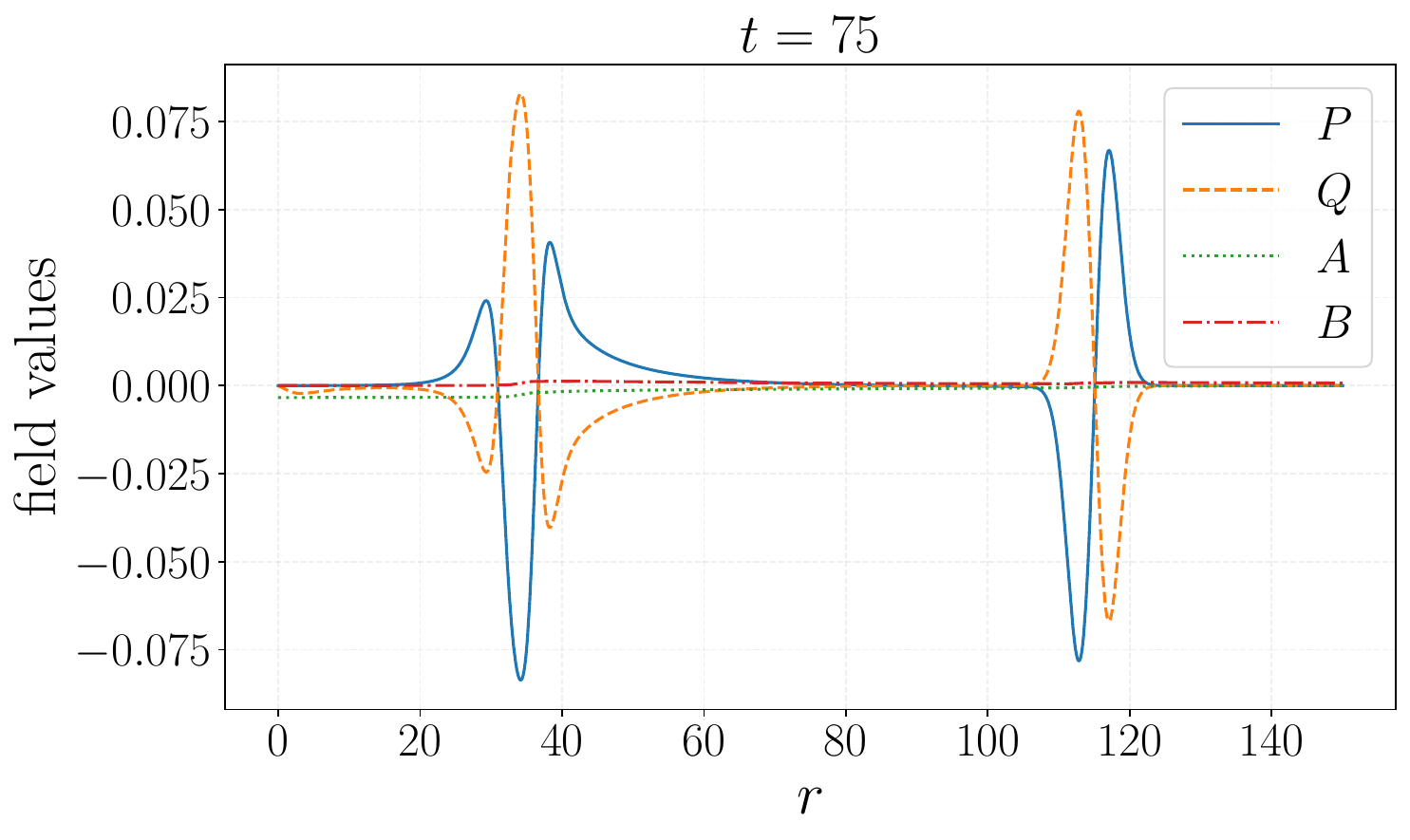} 
\includegraphics[scale=0.35]{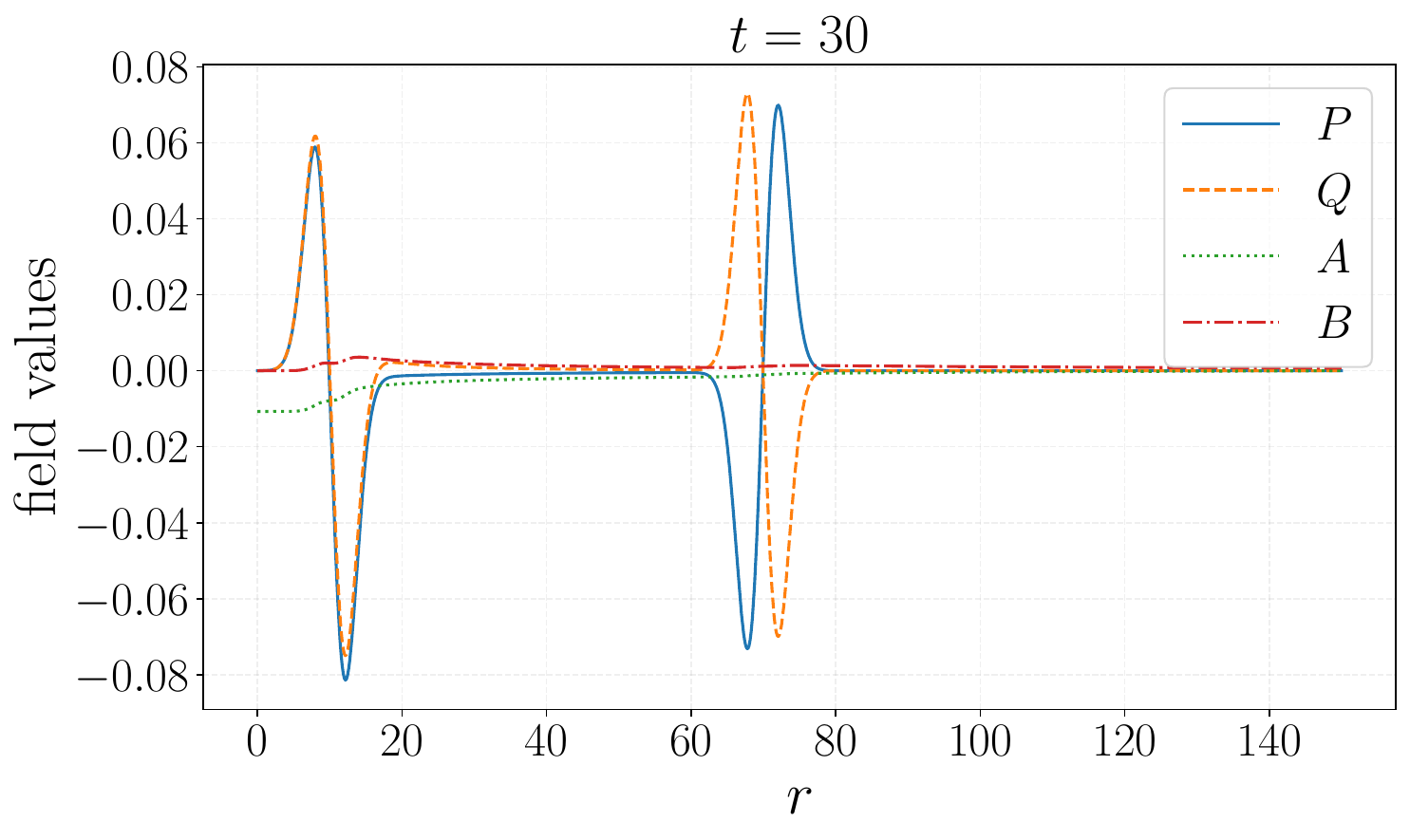}
\includegraphics[scale=0.35]{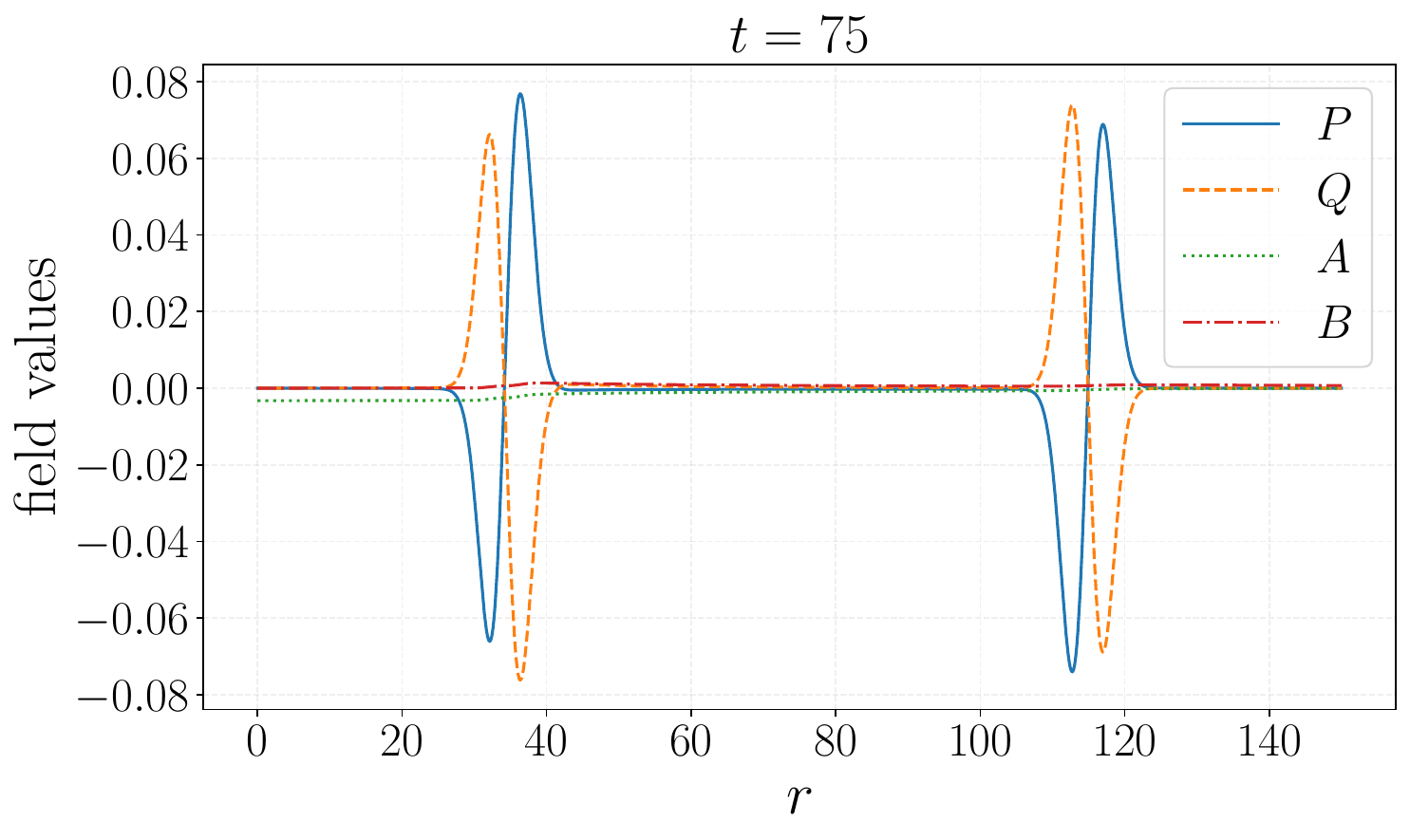} 
\includegraphics[scale=0.35]{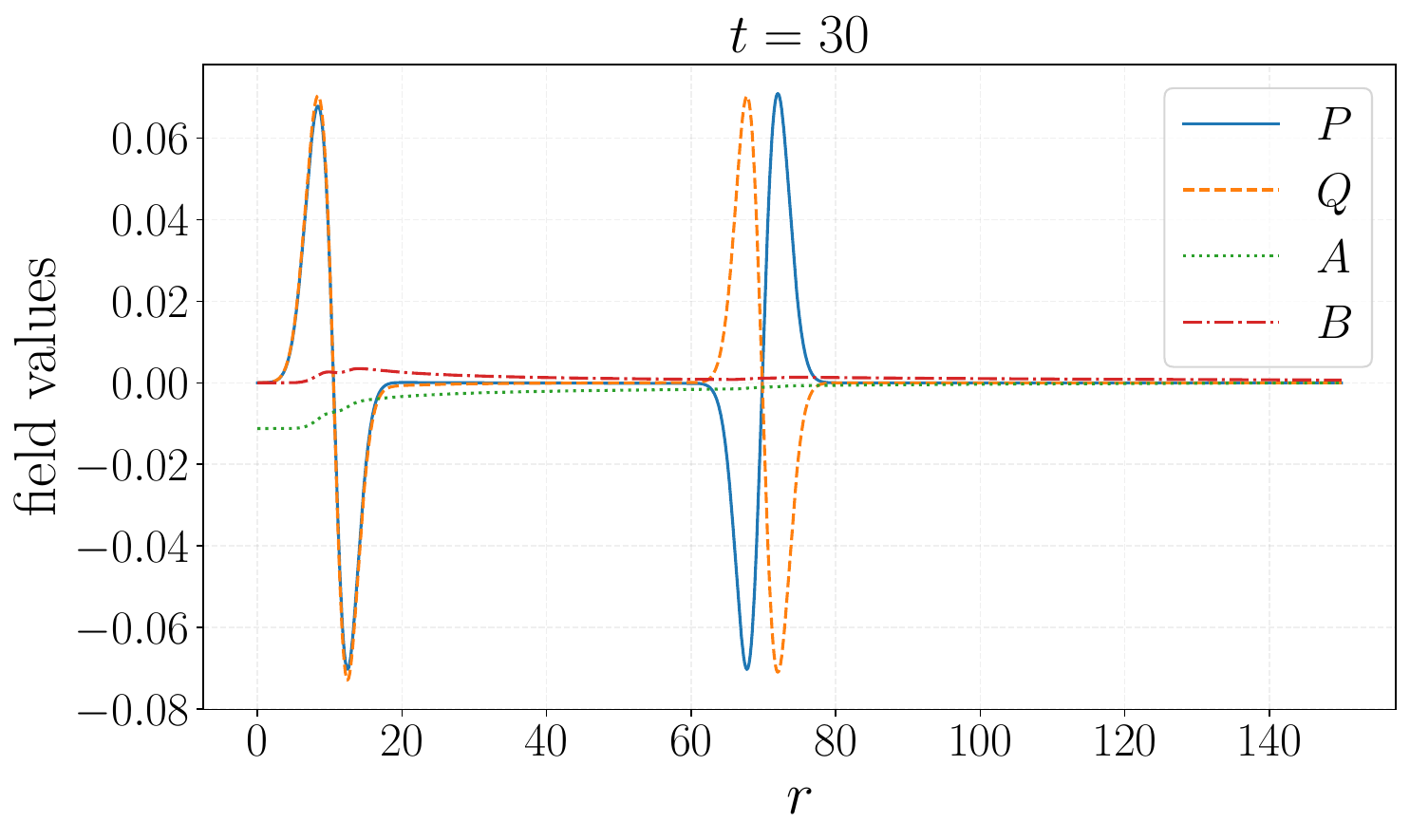} 
\includegraphics[scale=0.35]{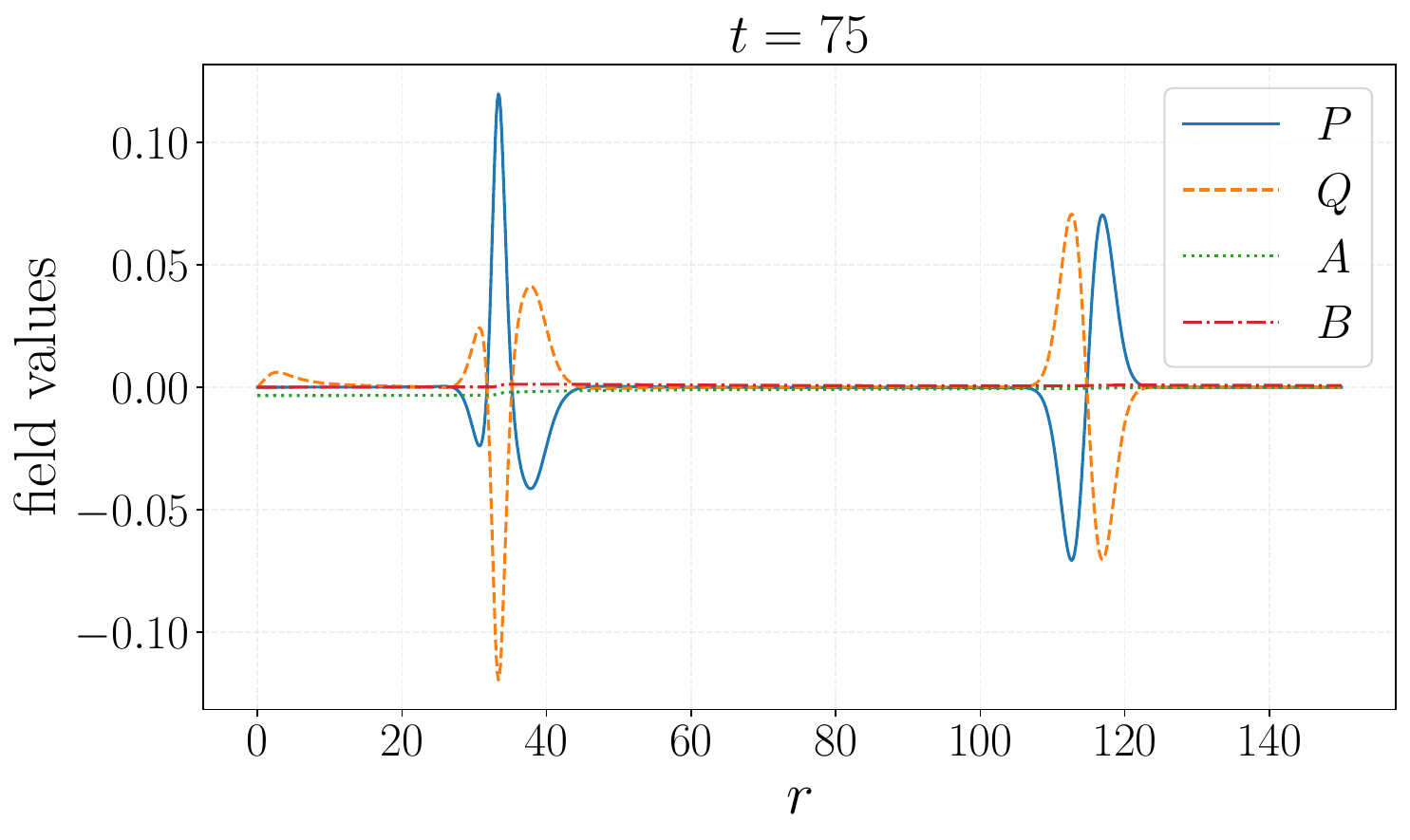}
\caption{\textbf{Type I data}: Snapshots of field variables for $\chi=-80$ \textit{(top panels)},  $\chi=0$ \textit{(middle panels)} and $\chi=50$ \textit{(bottom panels)}. Before the bounce (\textit{left panels}), the differences in the amplitudes of the fields are minimal across these three cases. However, after the ingoing pulse rebounds, shortly after $t>40$ (\textit{right panels}), the behavior of the field variables change appreciably. In particular, the case $\chi=50$ exhibits noticeably larger amplitudes compared to the other cases, highlighting the distinct dynamics induced by the self-interaction parameter.}
\label{fig:evol_variables} 	
\end{figure*}

\begin{figure*}[ht!]
\centering
 \includegraphics[scale=0.35]{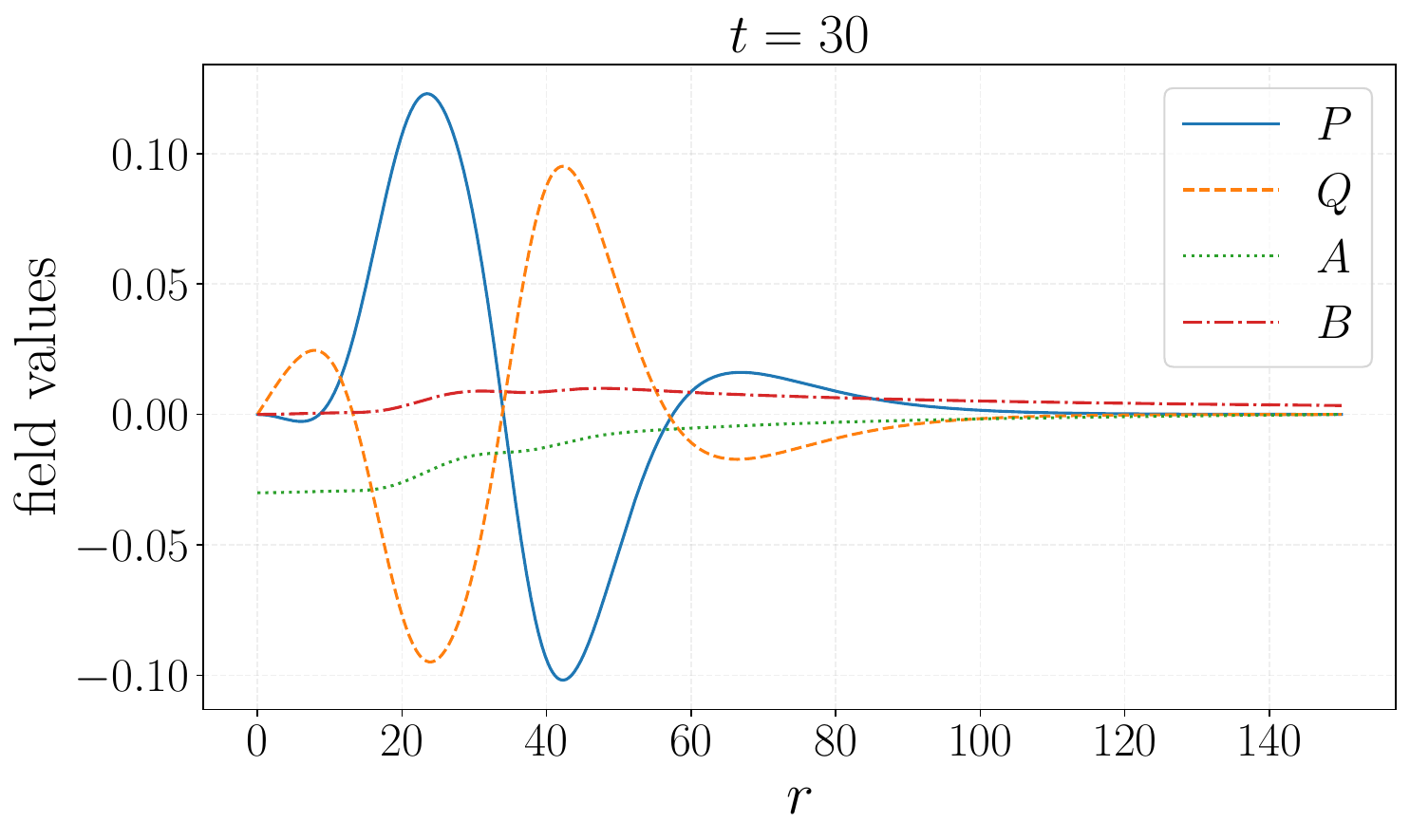}
\includegraphics[scale=0.35]{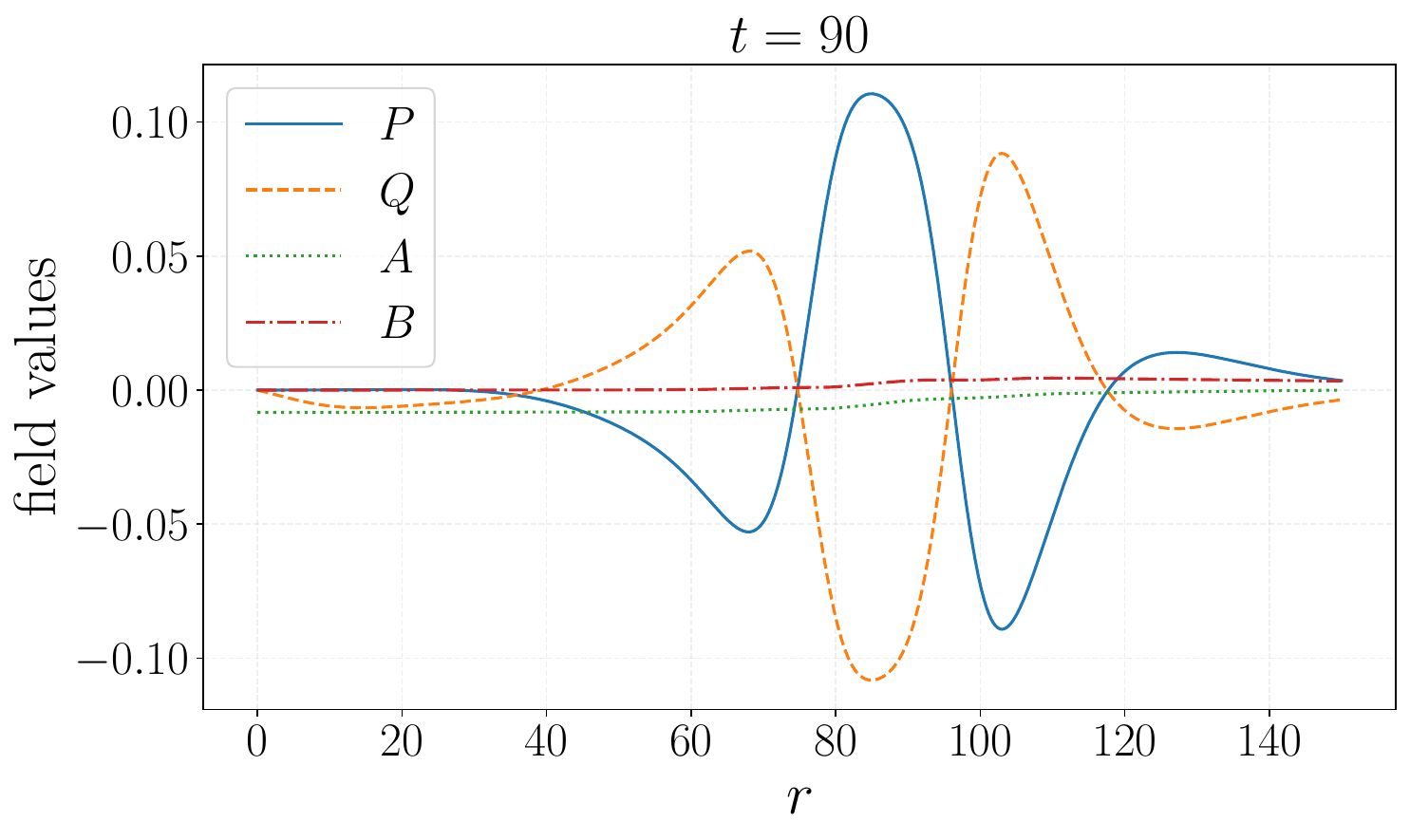} 
\includegraphics[scale=0.35]{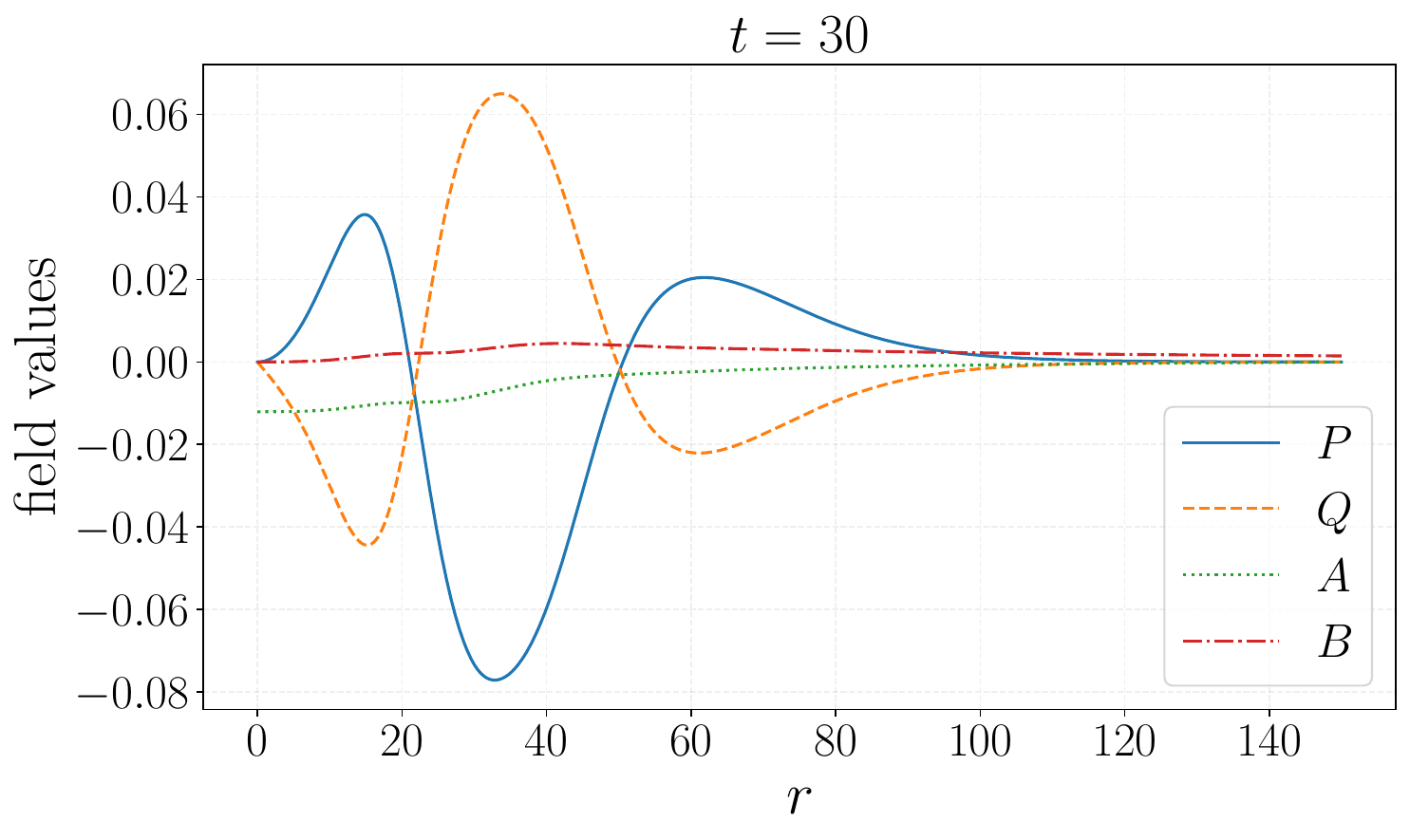}
\includegraphics[scale=0.35]{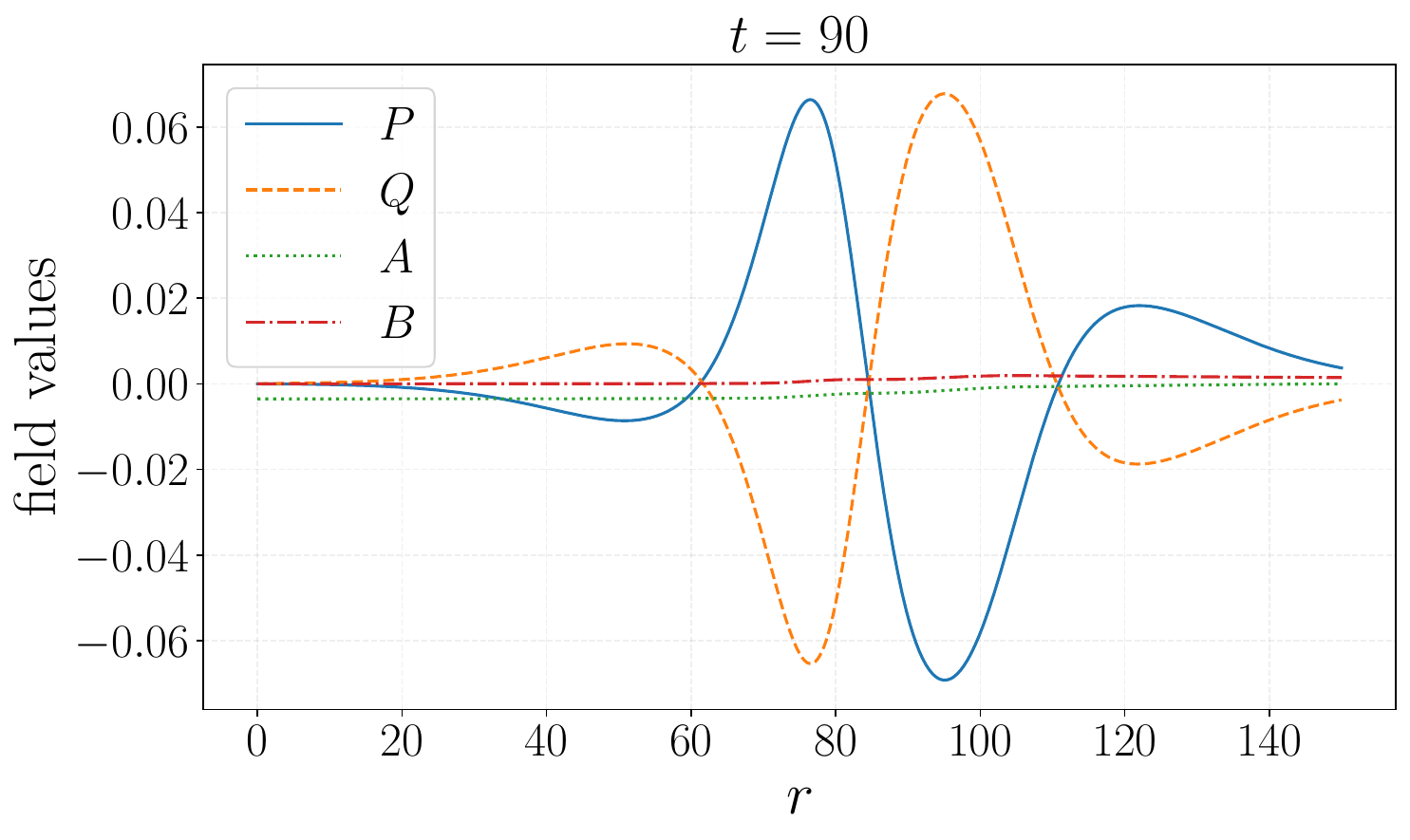} 
\includegraphics[scale=0.35]{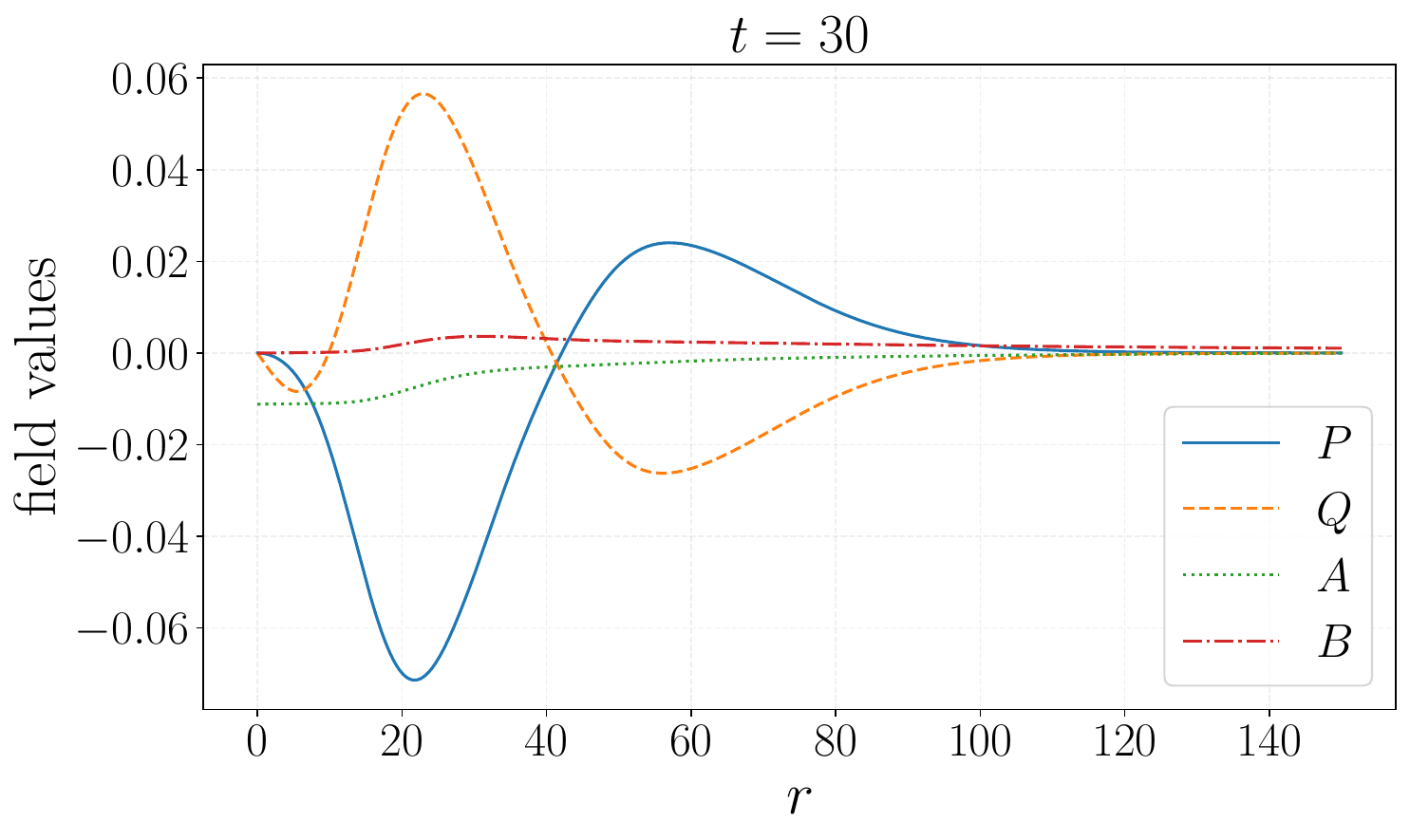} 
\includegraphics[scale=0.35]{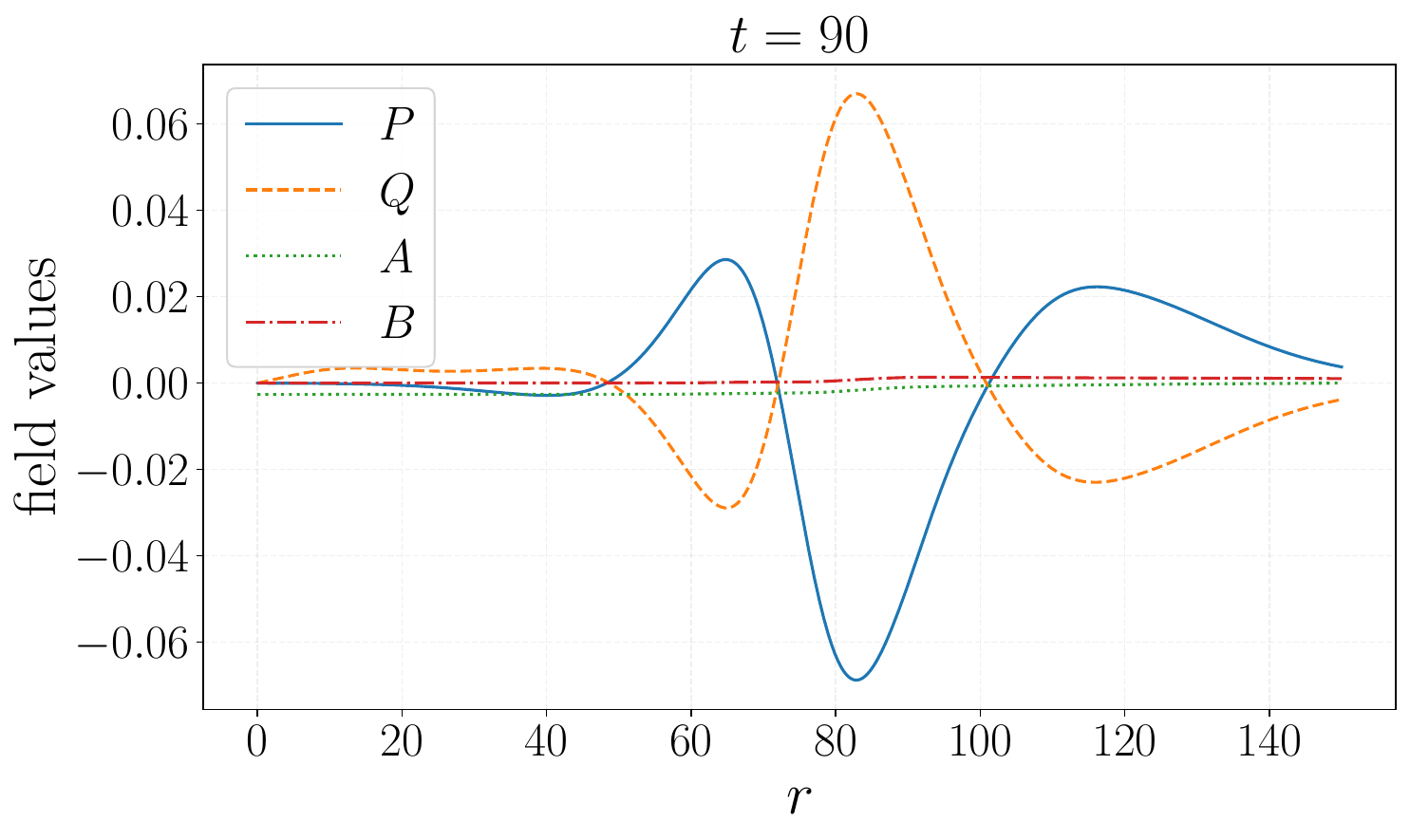}
\caption{\textbf{Type II data}: Snapshots of field variables for $\chi=-5$ \textit{(top panels)},  $\chi=0$ \textit{(middle panels)} and $\chi=1.2$ \textit{(bottom panels)}. There are noticeable differences in the amplitudes of the fields shortly after the onset of the simulation, significantly impacting the behavior of the field variables throughout the entire evolution. For a fixed time step (compare all three left panels with each other or all three right panels with each other), the effect of the self-interaction parameter on the dynamics is clearly evident.}
\label{fig:evol_variables2} 	
\end{figure*}

Finally, after analyzing the numerical evolution of the system across the entire computational domain, we find no evidence of hyperbolicity breakdown within the explored parameter space. This conclusion is supported by the well-behaved dynamics of the field variables, which exhibit smooth transitions and remain finite throughout the evolution.  For Type I data, Fig.~\ref{fig:evol_variables} illustrates three representative cases: \(\chi = -80\) \textit{(top panels)}, \(\chi = 0\) \textit{(middle panels)}, and \(\chi = 50\) \textit{(bottom panels)}. Similarly, for Type II data, Fig.~\ref{fig:evol_variables2} presents results for \(\chi = -5\) \textit{(top panels)}, \(\chi = 0\) \textit{(middle panels)}, and \(\chi = 1.2\) \textit{(bottom panels)}. We can confidently conclude that the self-interacting vector theory, characterized by an internal SU(2) global symmetry (Eq.~(\ref{eqn:action})), remains free of any pathological behavior even at the level of the fully non-linear dynamical regime.  
%

\subsection{A look towards spherical collapse}\label{sec:collapse}

Here, we study the spherical gravitational collapse within this theory considering various initial data\footnote{The study of critical collapse requires considering a general family of initial data \cite{Choptuik:1996yg,Maliborski:2017jyf}, as we do in this paper, since different types of critical behavior can occur. Thus, our analysis represents an important step toward a more comprehensive study of this critical phenomenon \cite{Gundlach:2007gc}.}. We refer an interested reader to Refs.\cite{Choptuik:1996yg,Choptuik:1999gh} for early studies on critical collapse of the Yang-Mill field and related studies \cite{Gundlach:1996je,Bizon:2010mp,Rinne:2013qc,Rinne:2014kka,Maliborski:2017jyf,Kain:2019jeg}.

While our current coordinate choice allows us to follow the system up to black hole formation, fully exploring the black hole's non-linear stability would benefit from employing horizon-penetrating coordinates or excising the region inside the apparent horizon. 

To begin, we increase the amplitude of the Type I data to $a_{0} = 0.535$, while keeping both $\chi = 50$ and $\mu=0.01$ fixed.  As a result, the leftward-moving pulse experiences a substantial increase in amplitude as it approaches $r=0$ (near the emergent horizon) without subsequent reflection. Meanwhile, for Type II data, we increase, instead, the self-interacting parameter from $\chi=1.2$ to $\chi=1.7$, keeping all the other parameters unchanged. A similar feature is captured here: the rightward-moving pulse experiences a significant increase in amplitude as it moves away from $r=0$.

Consequently, in both simulations, the increase in amplitude (for Type I data) and the self-interaction parameter (for Type II data) results in a significant enhancement of the energy density, leading to a strong gravitational attraction that prevents the pulse from fully dispersing. 
A key indicator of the formation of an apparent horizon is the vanishing expansion of the outgoing null geodesics (a marginally outer trapped surface) \cite{Thornburg:2006zb}. In spherical symmetry, this quantity is defined as $\theta\equiv e^{(A-B)}$. To confirm this, we have run multiple simulations, adjusting the parameters until this condition was met. 
\begin{figure*}[ht!]
\centering
\includegraphics[scale=0.35]{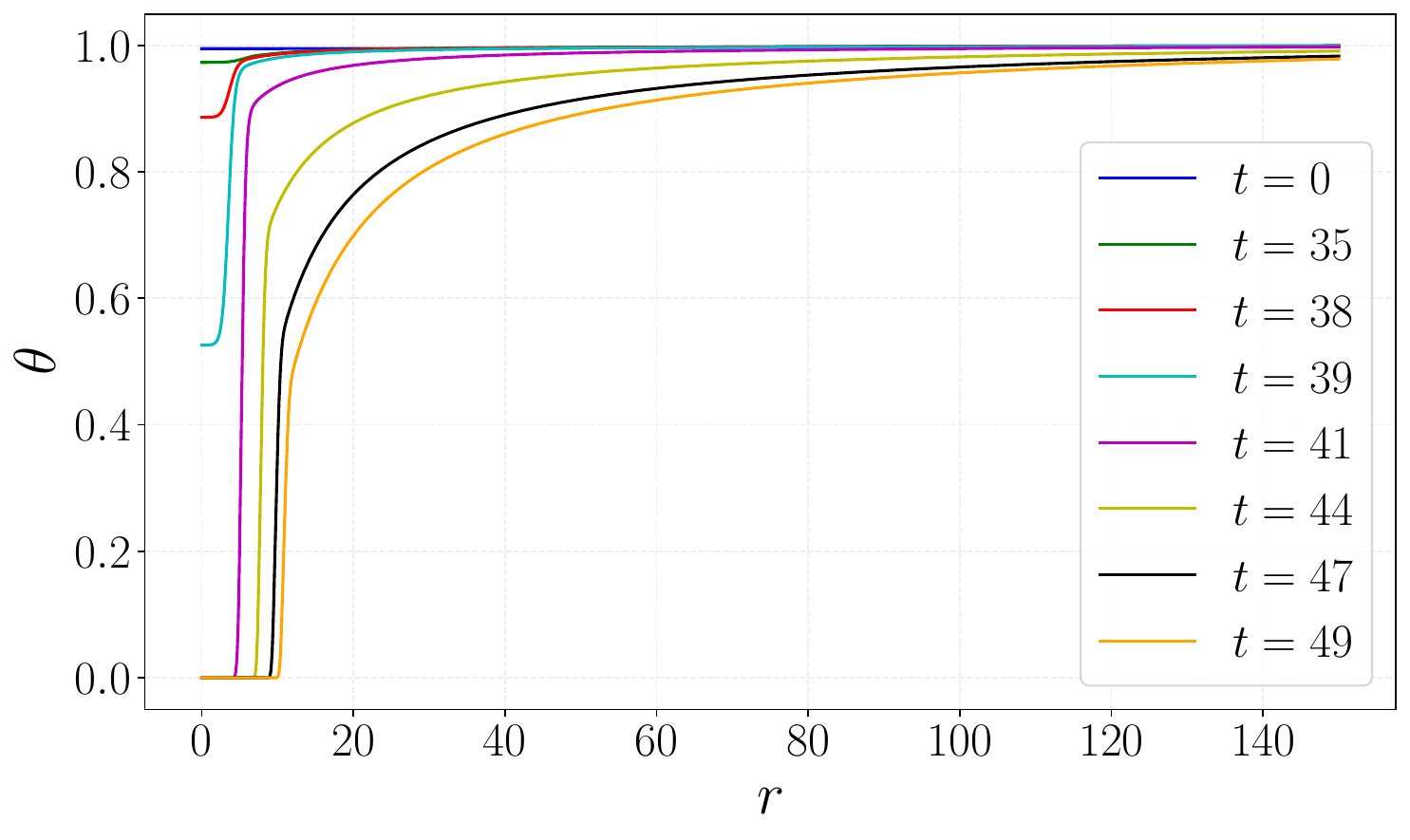}
\includegraphics[scale=0.35]{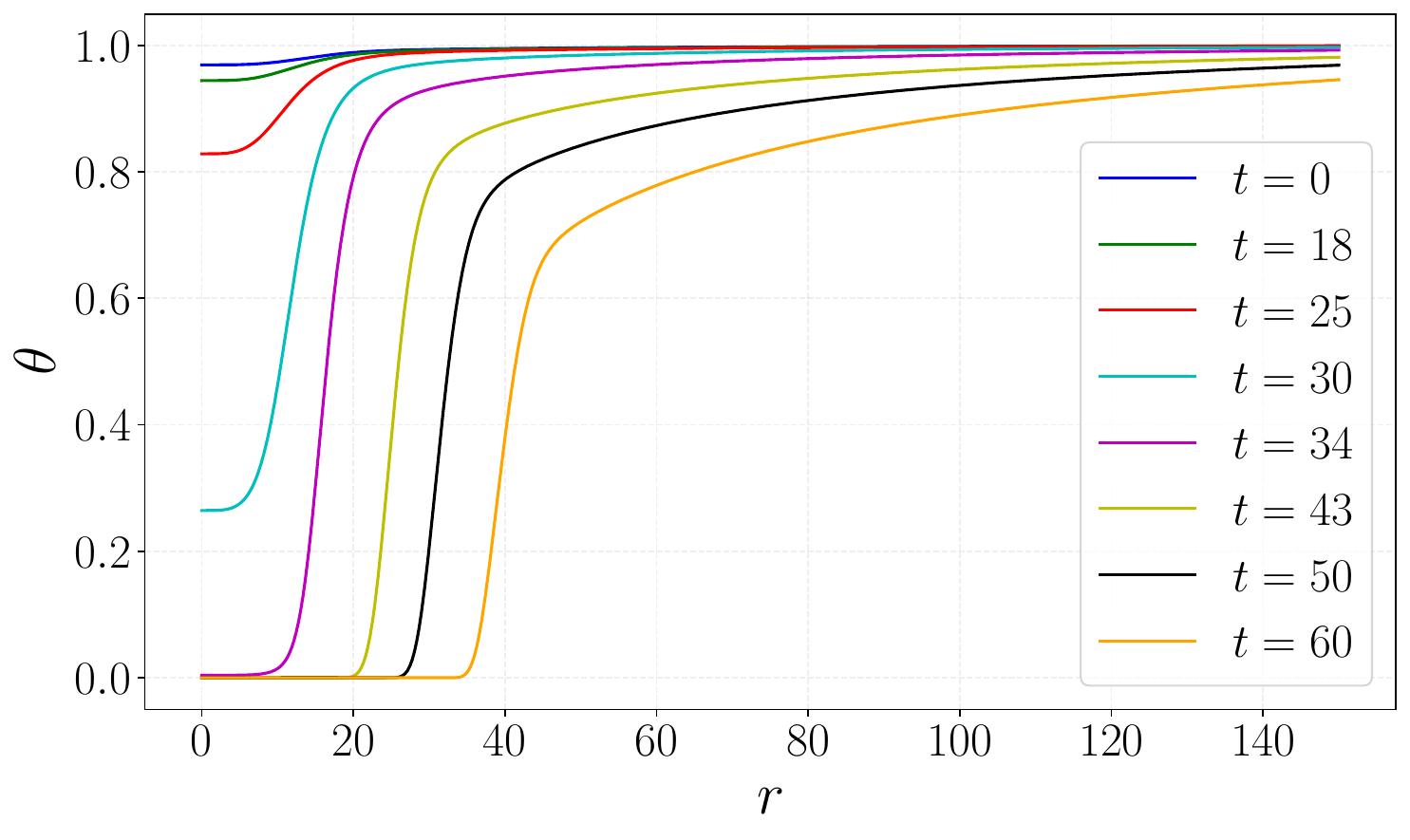}
\caption{Snapshots of the expansion of the outgoing null ray congruence. \textit{Left panel}: simulation of the  \textbf{type I data} with parameters $a_{0}=0.535$, $\mu=0.01$ and $\chi=50$.  \textit{Right panel}: simulation of the  \textbf{type II data} with parameters $a_{0}=-2.10479$, $r_{0}=30$, $w_{0}=20$, $\mu=0.01$, $b=0.104791$ and $\chi=1.7$. 
For Type I data, the expansion becomes confined at small radii due to enhancement of the energy density. In contrast, for Type II data, the expansion is more dispersed because of the outgoing nature of the pulse, leading to confinement around $r\approx40$, yet ultimately vanishing as it propagates through the numerical grid. This universal behavior, independent of the initial data, provides strong confirmation of the apparent horizon formation in this theory. 
}
\label{fig:expansion} 	
\end{figure*}

Fig.~\ref{fig:expansion} shows snapshots of the expansion up to the formation of the apparent horizon, capturing the spherical collapse of the ingoing pulse of the Type I data (left panel) and outgoing pulse of the Type II data (right panel). This occurs approximately at $t\gtrsim40$ for Type I data and $t\gtrsim35$ for Type II data, when $e^{(A-B)}$ approaches to zero, signaling the emergence of an apparent horizon. Notice, however, that for the Type II data, the vanishing of the expansion can occur even at larger points of the spatial grid ($r\gtrsim 10$) due to the nature of the profile.

Through an exploration of varying the amplitude and the self-interaction parameter $\chi$ for Type I data, we conclude that fixing the amplitude to, for instance, $a_{0}=0.535$, the collapse can be prevented for $\chi<50$ or induced for $\chi>50$. There are, of course, other parameter choices which lead to same conclusions. For instance, for Type I data, taking the mass $\mu=0.1$ and amplitude $a_{0}=0.96$ with the same $\chi=50$, the system also exhibits the formation of the apparent horizon. This highlights the diverse pathways to black hole formation, determined by multiple choices of parameter space and initial data values.

The formation of a black hole within this framework raises several intriguing questions that warrant further investigation. Notably, this aligns with the existence of an exact static charged black hole solution predicted by this theory in a former study \cite{Gomez:2023wei}, albeit in the massless case  $\mu=0$. This solution describes a Reissner-Nordstr\"{o}m-like black hole with a non-Abelian magnetic charge determined by the self-interaction parameter. However, the stability of this exotic solution remains unexplored. It should be thoroughly examined to determine whether spherical collapse can indeed lead to the formation of such a black hole or potentially reveal other undiscovered compact objects.

\subsection{Comparison with the Abelian case}

While we do not provide detailed calculations for the Abelian case, we highlight the key differences between the Abelian and non-Abelian cases, which lead to distinct features concerning the well-posedness of the Cauchy problem. One crucial point of distinction lies in the configuration of the vector field. In the Abelian case, it is well known that, under spherical symmetry, only the temporal and radial components of the vector field survive and must depend solely on the temporal and radial coordinates. As a result, the vector field is manifestly invariant under rotations—an invariance that excludes the possibility of the magnetic part. 

In contrast, in the SU(2) non-Abelian case, a more general configuration for the vector field is allowed within spherical symmetry, as described by Eq.~(\ref{eqn:witten}). 
A particularly relevant ansatz involves a purely magnetic gauge field configuration, which, as previously noted, has no analogue in the Abelian case. Hence, direct comparisons between the two cases can be misleading, as the Abelian configuration is purely longitudinal, whereas the non-Abelian configuration is purely transverse. Nevertheless, the structure of the wave equation in both cases can be examined to identify regimes where the system transitions from hyperbolic to parabolic or elliptic behavior. Such behavior arises only in the Abelian case. See Appendix \ref{sec:compariso}.

Another significant aspect concerns the Lorenz constraint. In the purely magnetic non-Abelian configuration, this condition is automatically satisfied without the need for additional constraints. By contrast, in the Abelian case, the Lorenz condition must be explicitly enforced at each point in the computational grid, modifying the principal part of the wave equation (see, e.g., \cite{Coates:2022qia}). This enforcement compromises the hyperbolic character of the field equations during nonlinear evolution. Technical analyses further show that the breakdown in well-posedness arises when the system transitions to a parabolic regime or when characteristic speeds become unbounded \cite{Rubio:2024ryv}. These pathologies are associated with mixed-type partial differential equations, such as Tricomi and Keldysh types \cite{Barausse:2022rvg,Ripley:2022cdh}.

Preliminary calculations for the dyon configuration (see Appendix~\ref{sec:Dyonmag}), which introduces an additional degree of freedom in the theory, suggest that similar dynamical shortcomings may emerge during nonlinear evolution, akin to those observed in the Abelian case. This highlights an important point: generality does not necessarily imply physical viability. Such extensions, while theoretically permissible, require dedicated follow-up studies within the present framework.

\section{Conclusions}\label{sec:conclusions}

In this work, we have presented compelling evidence that self-interacting vector fields naturally preserve their hyperbolic structure within the framework of the 't Hooft-Polyakov magnetic monopole configuration, which is inherently non-Abelian in nature. Although this configuration represents a specific ansatz within a broader class of configurations, It holds particular significance, as it gives rise to astrophysical solutions, such as gauge boson stars, black hole solutions, and neutron star solutions, as mentioned earlier. Therefore, the 't Hooft-Polyakov magnetic monopole configuration should not be regarded as a restricted case but rather as a physically meaningful and well-motivated solution, especially in contrast to more generic configurations, where self-gravitating solutions remain largely unexplored.

In contrast to the Abelian case, which faces significant challenges—such as instabilities and pathologies—ultimately leading to the breakdown of numerical simulations, we successfully evolve the Cauchy problem for self-interacting non-Abelian vector fields in spherical symmetry. Consequently, this formulation requires no corrective procedures on the equations of motion to restore hyperbolicity, such as \textit{fixing the equations} or invoking an ultraviolet completion for the Proca field. Notably, we have observed stable numerical evolutions of the vector field throughout the entire time domain and across a broad range of initial conditions within a fully dynamical background, providing strong evidence for the well-posedness of the IVP in this theory. This outcome stands as a clear counterexample to previous claims about the inherent instability of self-interacting vector fields. \textit{In this context, the celebrated problems associated with self-interacting Proca fields do not manifest in the case of SU(2) Proca fields.} 

Our simulations have also provided valuable insights into the physics of wave propagation in this framework, particularly in how the self-interaction parameter governs the nonlinear dynamics of the vector field. Interestingly, the results reveal regimes where dispersion, reflection, and amplification either emerge or remain absent during the system's evolution. For type I data, we have observed an oscillatory behavior in the ingoing wave packet as it undergoes partial dispersion and reflection. This effect becomes apparent when the particle mass increases to $\mu\sim \mathcal{O}(0.1)$. We identify this as the large-mass limit, where the oscillatory frequency scales as $\omega\sim \mu$. While this behavior is not the primary focus of our study, it may warrant further investigation in the context of spherical collapse.

Finally, understanding black hole formation in this framework will not only validate the existence of the static charged black hole solution predicted by this theory but also shed light on its stability and long-term dynamics. Investigating the dynamics of this solution could provide valuable insights into the interplay between self-interacting vector fields and gravitational collapse, potentially uncovering new phenomena intrinsic to non-Abelian theories. 

For future work, we plan to investigate both the hyperbolic character and spherical collapse within a more general vector field profile that incorporates the electric component of the gauge field, known as the \textit{dyon}. We conjecture that, unlike the 't Hooft-Polyakov magnetic monopole, there may exist regions of spacetime where the character of the PDEs governing the vector field dynamics changes from hyperbolic to elliptic. In this context, it will be crucial to identify, if they exist, regions of the parameter space that lead to the loss of hyperbolicity outside the horizon. To address these challenges, we must employ horizon-penetrating coordinates, which would not only enable the study of black hole formation and stability but also allow for an exploration of ``hairy'' black hole scenarios with non-trivial vector field profiles. Such findings may have significant implications for future gravitational wave experiments, particularly in the context of black hole binary evolutions in extreme environments—i.e., beyond the traditional vacuum assumption— and contribute to the broader comprehension of strong-field gravity in extreme environments.  

\section*{Acknowledgments}

Special thanks go to Nicola Franchini for sharing the numerical code. We are deeply grateful to Miguel Bezares for providing invaluable guidance during the numerical implementation. We also thank Miguel for carefully reviewing this manuscript and offering insightful critiques and comments, which have significantly contributed to refining this final version. Finally, we acknowledge Fethi Ramazanoglu for his insightful remarks on the implications of our results. J. F. R. is thankful for financial support
from the Universidad Industrial de Santander, VIE, postdoctoral Contract No. N°003-4512 of 2025 with contractual registry No. 2025000174.

\appendix
\section{Noether currents}\label{append:A}
The action \eqref{eqn:action} is invariant under internal global SU(2) transformations, which are given by
\begin{equation}
    B_{a\mu} \mapsto B_{a\mu} + \delta B_{a\mu} = B_{a\mu} - i \epsilon_{abc} \lambda^b A^{c}{}_\mu, \label{eqn:globalsu2}
\end{equation}
where $\lambda^{b}$ are constants. Under these global transformations, the Lagrangian changes as follows:
\begin{align}
    \delta L &= \frac{\delta L}{\delta B_{a\mu}} \delta B_{a\mu} + \frac{\delta L}{\delta \nabla_\nu B_{a\mu}} \delta \nabla_\nu B_{a\mu}, \\
    &= \nabla_\nu \left[ \frac{\delta L}{\delta \nabla_\nu B_{a\mu}} \delta B_{a\mu} \right], \\
    &= \nabla_\nu J^\nu = 0,
\end{align}
where we have used the field equations, and $J^{a\mu}$ represents three currents given by
\begin{equation}
    J^{a\nu} \equiv \frac{\delta L}{\delta \nabla_\nu B_{b\mu}} \epsilon^{a}{}_{bc} B^c{}_{\mu}.
\end{equation}
Since the Lagrangian is invariant under \eqref{eqn:globalsu2}, the currents $J^{a\nu}$ are conserved. 

More precisely, for the case under study, the conserved currents are
\begin{equation}
    J^{a\nu} = F^{b\nu\mu} \epsilon^{a}{}_{bc} B^c{}_{\mu}.
\end{equation}
From these results, we can obtain the following conserved charges:
\begin{equation}
    Q^a = \oint F^{b0\mu} \epsilon^{a}{}_{bc} B^c{}_{\mu} \sqrt{-g} \, d^3 y.
\end{equation}
For the magnetic monopole configuration, these charges are all zero. However, for more general profiles, they do not vanish and might be included as additional consistency constraints.

\section{Dyon and magnectic monopole configurations}\label{sec:Dyonmag}
From the most general configuration, we can derive two distinct cases. The first configuration is characterized by $A_1 = \phi_1 = 0$ and $A_0 \neq 0$, $\phi_2 \equiv w \neq 0$. When $A_0 = 0$, the Dyon reduces to the 't Hooft-Polyakov monopole. The constraint derived from the field equations \eqref{eqn:lorentzgen} for the Dyon case is given by:
\begin{multline}
    A_0 e^{2A} \mu^2 r^2 \left(\dot{A} - \dot{B}\right) + r^2 \left(\chi_1 + \chi_2\right) A_0^3 \left(3 \dot{A} - \dot{B}\right) \\
    - 2 A_0 e^{2A} (w + 1) \chi_1 \left[(w + 1) \left(\dot{A} - \dot{B}\right) - 2 \dot{w}\right] = 0.
\end{multline}
As can be seen, this constraint \emph{is automatically satisfied for the magnetic monopole case}, $A_0 = 0$.

\section{Comparison between the Abelian and non-Abelian cases}\label{sec:compariso}

As already mentioned, the key difference stems from the generalized Lorentz gauge condition, which is not automatically satisfied in the Abelian case. As a result, this imposes constraints on the initial conditions. Moreover, the generalized Lorentz gauge modifies the principal symbol. In contrast, for the non-Abelian pure magnetic monopole ansatz, no such constraint on the initial conditions arises, meaning that the principal symbol remains unmodified. To illustrate this point, we perform the analysis on flat spacetime for both models with a quartic self-interaction.

The Abelian field will be denoted by $A_\mu$, and the field equations are given by
\begin{equation}
    \nabla_\mu F^{\mu\nu} = z \mu^2 A^\nu, \label{eqn:FEA}
\end{equation}
where $F_{\mu\nu} = \nabla_\mu A_\nu - \nabla_\nu A_\mu$ and $z = (1 + \lambda A_\alpha A^\alpha)$. 

The generalized Lorentz gauge is
\begin{equation}
    \nabla_\mu (z A^\mu) = 0. \label{eqn:lorentzA}
\end{equation}
This expression is a consequence of the field equations, so it does not represent a gauge choice despite its name. It can be used to derive an alternative form of the field equations  \cite{Coates:2022qia, Coates:2023dmz}:
\begin{multline}
    g_{\alpha\beta}^{\rm eff} \nabla^\alpha \nabla^\beta A_\nu + 2 \lambda A^\mu A^\alpha \nabla_\mu F_{\nu\alpha} \\
    + 2 z \lambda \nabla_\mu A_\alpha \nabla_\nu (z^{-1} A^\mu A^\alpha) - z^2 \mu^2 A_\nu = 0, \label{eqn:manifestly}
\end{multline}
where $g_{\alpha\beta}^{\rm eff}=z g_{\alpha\beta} + 2 \lambda A_\alpha A_\beta$ is the effective metric.
The field equations \eqref{eqn:FEA} are equivalent to \eqref{eqn:manifestly} together with the generalized Lorentz gauge \eqref{eqn:lorentzA}. For the 1+1 case, the highest derivatives come only from the term proportional to the effective metric. The presence of the disformal term in the effective metric can cause a loss of hyperbolicity, which is the central issue we aim to highlight here.

In the 1+1 spherically symmetric configuration that involves only the radial component $A_\mu = (0, r^{-1} v, 0, 0)$, the determinant of the effective metric is
\begin{equation}
    \det\Bigl(g_{\alpha\beta}^{\rm eff}\Bigr)=-\frac{\left(r^2+\lambda v^2\right) \left(r^2+3 \lambda v^2\right)}{r^4},
\end{equation}
and it can vanish when $\lambda < 0$, more exactly, when the field reaches the values $v_{1,2} = \pm r/(-\lambda)^{1/2}, \pm r/(-3\lambda)^{1/2}$. Therefore, self-interaction can induce hyperbolicity loss. To make this more explicit, the dynamical field equation in this 1+1 case reduces to
\begin{multline}
    -(1 + \lambda v^2 / r^2) \partial_t^2 v + (1 + 3 \lambda v^2 / r^2) \partial_r^2 v \\
    + \frac{4 \lambda v (\partial_r v)^2}{\lambda v^2 + r^2} - \frac{4 \lambda v^2 (\lambda v^2 + 3 r^2) \partial_r v}{r^3 (\lambda v^2 + r^2)} \\
    + \frac{2 (\lambda^2 v^5 + 2 \lambda r^2 v^3 - r^4 v)}{r^4 (\lambda v^2 + r^2)} - \frac{\mu^2 v (\lambda v^2 + r^2)^2}{r^4} = 0,
\end{multline}
where two distinctive features arising from self-interaction become evident: a modification of the principal part of the equations and nonlinearities. The former is especially critical: a wrong sign in the principal part can spoil the hyperbolic character of the system, causing frequency modes to grow exponentially over time. As a result, even small initial perturbations may evolve into large amplitudes, resulting in a breakdown of predictivity.

In contrast, for the non-Abelian case under the 't Hooft–Polyakov configuration, the field equation takes the form
\begin{multline}
    -\partial_t^2 w + \partial_r^2 w - \frac{w (w^2 - 1)}{r^2} 
    - \mu^2 (w + 1) +\frac{\chi (w + 1)^3}{r^2} = 0.
\end{multline}
This equation remains hyperbolic regardless of the value of the self-interaction parameter $\chi$, and no additional constraints on the initial data are required.

\section{Convergence test}\label{append:C}
\begin{figure*}[ht!]
\centering
\includegraphics[scale=0.35]{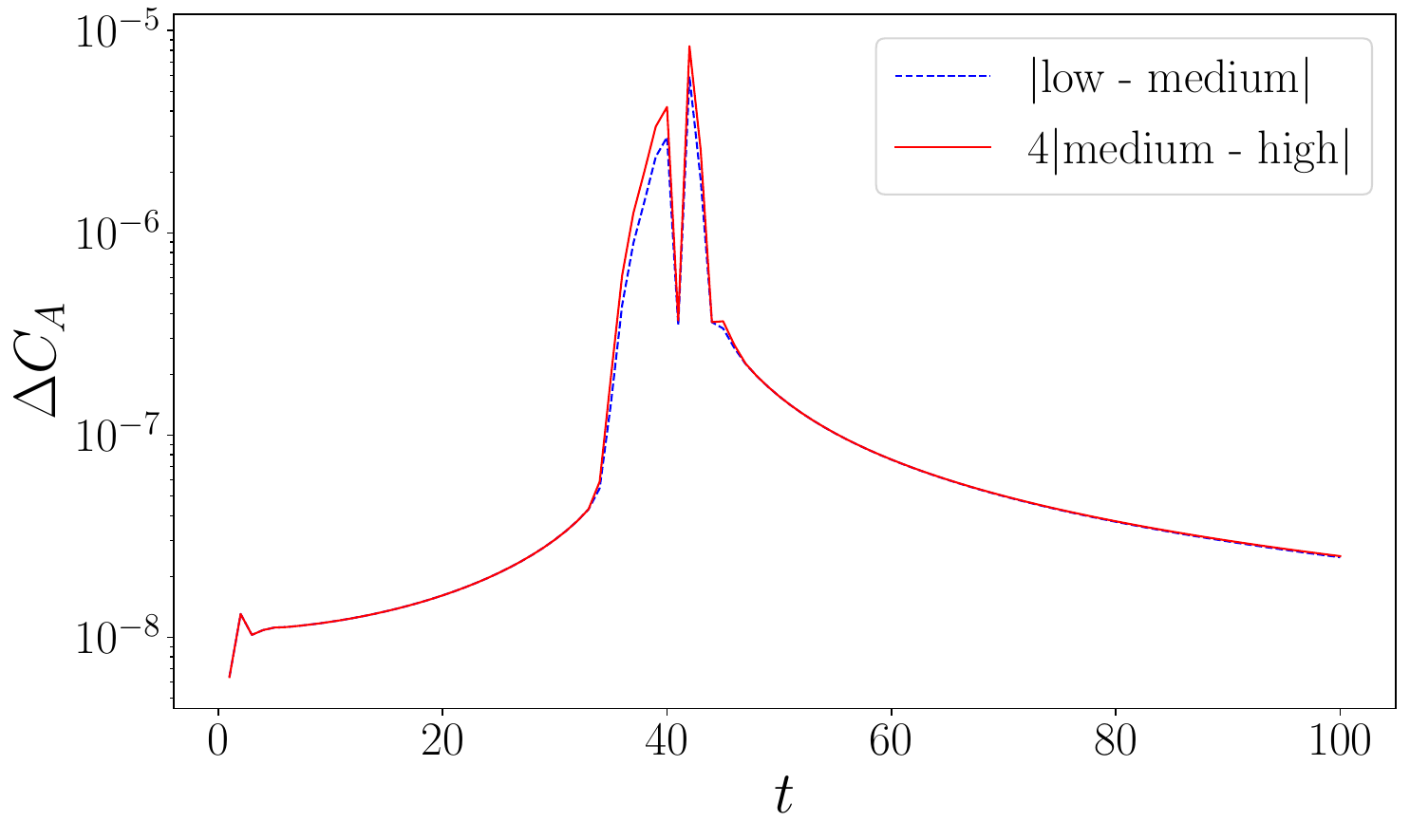} 
\includegraphics[scale=0.35]{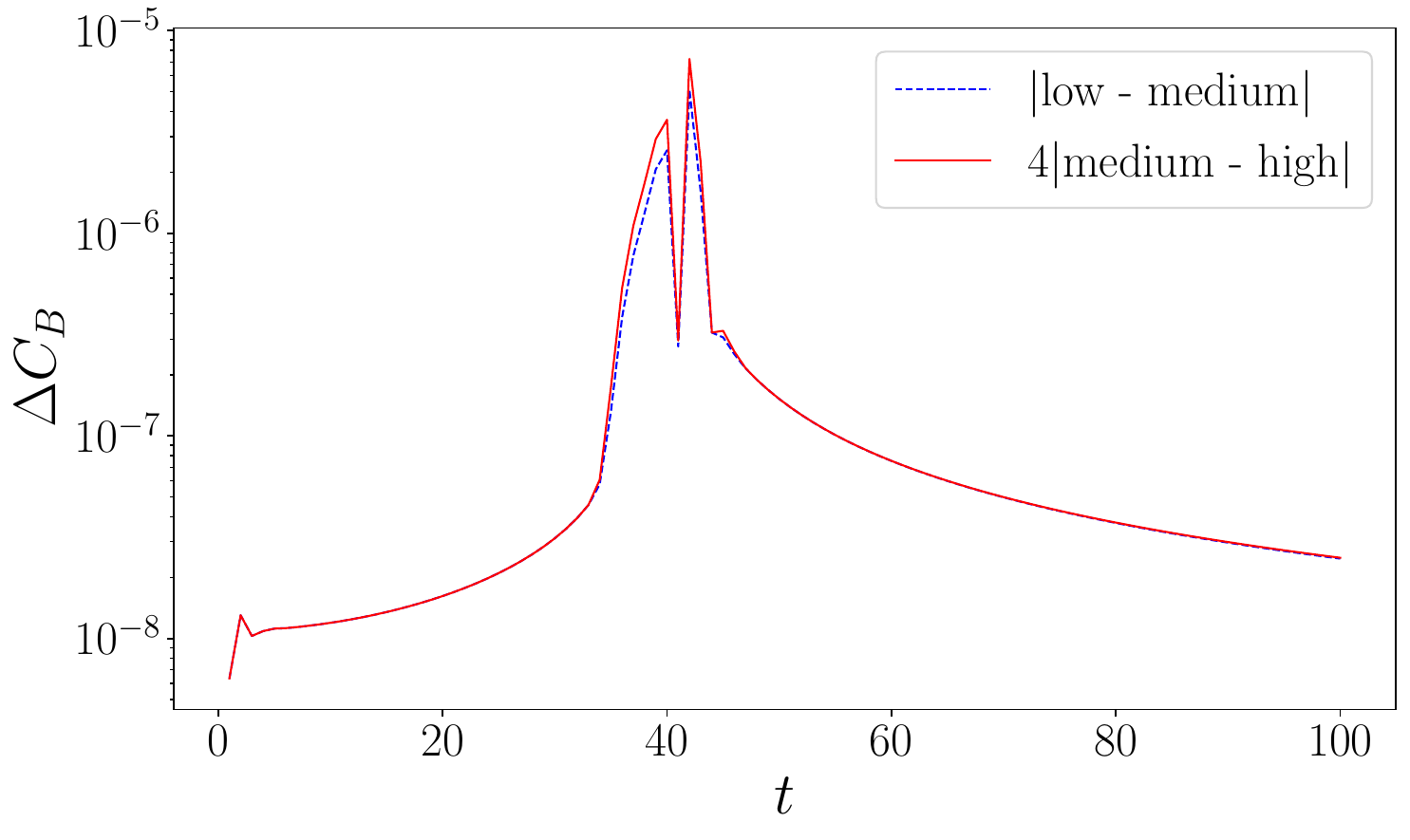} 
\caption{Convergence of the constraint equations $C_{A}$ and $C_{B}$ in the left and right panels, respectively. We have computed the absolute difference between the low and medium resolutions (dashed blue line) and between the medium and high resolutions (dotted red line). The latter is scaled by a factor of $c_{2} = 4$, confirming second-order convergence.}
\label{fig:convertest} 	
\end{figure*}

To check the expected second-order convergence of the numerical solution, we perform a standard test by computing the convergence factor associated with the truncation error. This is done using three different spatial resolutions: $\Delta x_{\rm low}=0.01$,\;$\Delta x_{\rm med}=0.005$\; and $\Delta x_{\rm high}=0.0025$. This factor is defined as \cite{10.5555/1403886}
\begin{equation}
    c_{p}=\frac{|(\Delta x_{\rm low})^{p}-(\Delta x_{\rm mid})^{p}|}{|(\Delta x_{\rm mid})^{p}-(\Delta x_{\rm high})^{p}|},\label{eqn:convfactor}
\end{equation}
which scales, in the continuum, as $c_{p}\approx 2^{p}$. In Fig.~\ref{fig:convertest}, the convergence factor of the constraint equations $C_{A}$ and $C_{B}$ are displayed, showing a second order convergence as expected.

\begin{figure*}[ht!]
\centering
\includegraphics[scale=0.45]{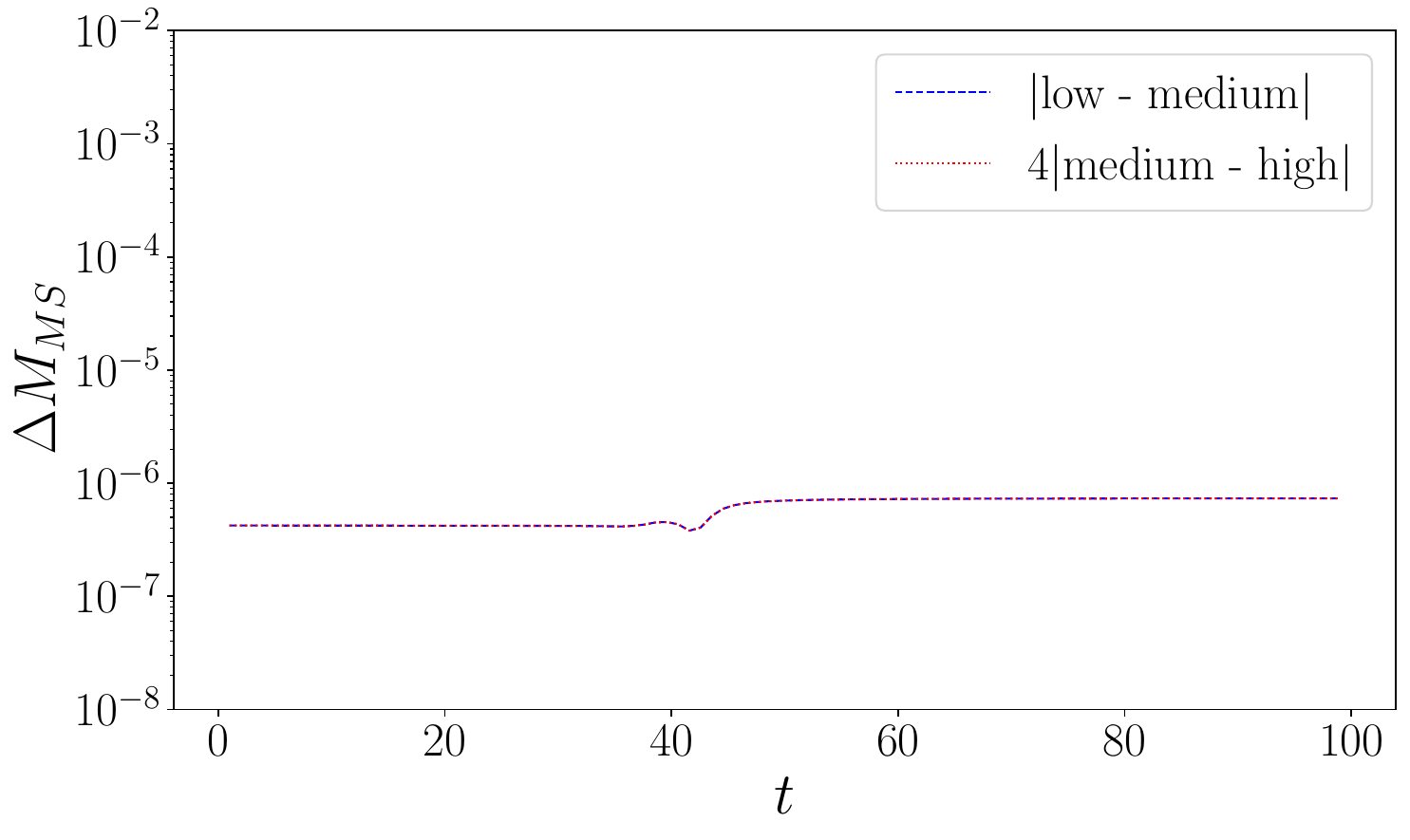} 
\caption{Convergence for the Misner-Sharp mass. This result displays the absolute difference between the low and medium resolutions (dashed blue line) and the medium and high resolutions (dotted red line). The latter is rescaled by a factor $c_{2} = 4$, showing, as expected, second order convergence.}
\label{fig:Misner_mass} 	
\end{figure*}
In addition, we present in Fig.~\ref{fig:Misner_mass} the absolute difference of the Misner-Sharp mass to further verify convergence in a conserved quantity. In spherical symmetry, and considering the line element given by Eq.~(\ref{eqn:line_element}), we calculate the Misner-Sharp mass as follows \cite{Misner:1964je,Hayward:1994bu}
\begin{equation}
    M_{\rm MS}(t,r)=\left.\frac{r}{2}(1-e^{-2B(t,r)})\right|_{r=r_{\rm max}}.
\end{equation}
In doing so, we extract the value of the mass at the last point of the spatial grid (which represents spatial infinity in our simulations) for the specified resolutions, taking care to account for boundary effects. As expected, this quantity remains constant 
throughout the time evolution, reinforcing the reliability of our numerical implementation. More importantly, this analysis confirms a second-order convergence as the resolution increases, demonstrating both the consistency and convergence of the numerical solutions.

\bibliography{biblio}

\end{document}